\documentclass[twocolumn,notitlepage,superscriptaddress, nofootinbib,nobibnotes, aps,prd,10pt]{revtex4-1}
\usepackage{amsmath,amssymb, fp}  
\usepackage{xcolor}  
\usepackage{soul} 
\definecolor{linkcolor}{RGB}{150,50,60}  
\usepackage[	colorlinks=true,    	
			pdfstartview=FitV,
			linkcolor= linkcolor,
			citecolor= linkcolor,
			urlcolor= linkcolor,
			linktoc=page,
			hyperindex=true,
			hyperfigures=true,  
			breaklinks=true]
			{hyperref}
\usepackage{dblfloatfix}    
\usepackage{balance} 
\usepackage{lastpage}  
\usepackage{enumitem}
\usepackage{mathtools}
\usepackage{epsfig,rotating,pifont}
\usepackage{graphicx}
\graphicspath{{./figures/}}	
\usepackage[caption=false]{subfig}
\usepackage{placeins}

\usepackage{ragged2e} 
\usepackage{etoolbox}
\usepackage{changepage}   
\usepackage{pgfplots}
\usetikzlibrary{pgfplots.groupplots}
\pgfplotsset{compat=1.3}
\usepackage{tikz}  
\usetikzlibrary{arrows,shapes,positioning}
\usetikzlibrary{decorations.markings,decorations.pathmorphing,decorations.pathreplacing,decorations.text}
\usetikzlibrary{arrows,calc,patterns,shapes.geometric}
\usepackage{fancybox}
\usepackage{verbatim}
\usepackage{multirow}
\usepackage{titlesec}

\newsavebox\CBox 
\def\textBF#1{\sbox\CBox{#1}\resizebox{1\wd\CBox}{0.97\ht\CBox}{\textbf{#1}}}

\def\beq{\begin{eqnarray}}
\def\eeq{\end{eqnarray}}

\def\a{\alpha}

\def\eps{\epsilon}

\def\la{\langle }
\def\ra{\rangle }

\def\lb{\label}

\def\M{\mathcal{M}}
\def\H{\mathcal{H}}

\def\B{\mathcal{B}}

\def\E{\mathcal{E}}

\newcommand{\Tr}{\mathrm{Tr}\hspace{1pt}}            

\newcommand{\be}{\begin{equation}}
\newcommand{\ee}{\end{equation}}
\newcommand{\bea}{\begin{eqnarray}}
\newcommand{\eea}{\end{eqnarray}}
\newcommand{\bg}{\begin{gather}}

\newcommand{\bseq}{\begin{subequations}}
\newcommand{\eseq}{\end{subequations}}

\newcommand{\bra}[1]{\langle #1 |}
\newcommand{\ket}[1]{| #1 \rangle}

\def\tr{{\rm Tr}}

\def\be{\begin{eqnarray}}
\def\ee{\end{eqnarray}}
\def\lb{\label}

\def\dra{{\ra\hspace{-2pt}\ra}}
\def\dla{{\la\hspace{-2pt}\la}}
\def\n{\mathrm{n}}
\def\h{\mathfrak{h}}
\def\S{\mathcal{S}}
\def\nn{\nonumber}

\definecolor{Green}{RGB}{147,162,153}
\definecolor{Green2}{RGB}{26,148,49}
\definecolor{BrownL}{RGB}{173,143,103}
\definecolor{Red}{RGB}{210,83,60}
\definecolor{BrownD}{RGB}{114,96,86}
\definecolor{GreyD}{RGB}{76,90,106}
\definecolor{GreyB}{RGB}{128,141,160}
\definecolor{Maroon}{RGB}{121,70,61}
\definecolor{Blue}{RGB}{148,184,210}
\definecolor{Blue2}{RGB}{108,144,170}
\definecolor{Blue3}{RGB}{42, 107, 172}
\definecolor{BB}{RGB}{128,184,220}

\newsavebox\foobox
\begin{document}

\title{Topological reflected entropy in Chern-Simons theories}  

\author{Cl\'ement Berthiere}  
\email{clement.berthiere@pku.edu.cn}

\author{Hongjie Chen}  
\email{chenhongjie01@pku.edu.cn}

\author{Yuefeng Liu}  
\email{yfliu0905@pku.edu.cn}

\address{School of Physics, Peking University, Beijing 100871, China}  

\author{Bin Chen}  
\email{bchen01@pku.edu.cn}

\address{School of Physics, Peking University, Beijing 100871, China}  
\address{Collaborative Innovation Center of Quantum Matter, Beijing 100871, China}
\address{Center for High Energy Physics, Peking University, Beijing
100871, China}

\date{\today} 
 
\begin{abstract}\vspace{4pt}
\begin{center}\textbf{\abstractname}\end{center}\vspace{-5pt} 
We study the reflected entropy between two spatial regions in $(2+1)$--dimensional Chern-Simons theories. Taking advantage of its replica trick formulation, the reflected entropy is computed using the edge theory approach and the surgery method. Both approaches yield identical results.  In all cases considered in this paper, we find that the reflected entropy coincides with the mutual information, even though their R\'enyi versions differ in general. We also compute the odd entropy with the edge theory method. The reflected entropy and the odd entropy both possess a simple holographic dual interpretation in terms of entanglement wedge cross-section. We show that in $(2+1)$--dimensional Chern-Simons theories, both quantities are related in a similar manner as in two-dimensional holographic conformal field theories (CFTs), up to a classical Shannon piece.

\end{abstract}

\maketitle  

\makeatletter
\makeatother
\tableofcontents

\section{Introduction} \label{sec:intro}

Quantum information has established new perspectives to investigate various areas of physics, such as \mbox{quantum} field theory, condensed matter physics, and quantum gravity. Central to these developments is the concept of quantum entanglement, which has proven to be a formidable tool to characterize quantum many-body systems. This is particularly true for topological states of matter which cannot be identified via conventional local order parameters or correlations functions. The topological entanglement entropy \cite{Kitaev:2005dm,Levin:2006zz} encodes information about the topological order of ground states of gapped systems. This quantity arises as a universal finite contribution in the entanglement entropy of two-dimensional spatial subregions for such states.

The entanglement entropy is, however, only a proper measure of entanglement for pure quantum states, and does not give a meaningful picture of correlations for more general states. Other information-theoretic quantities then have to be considered for mixed states, and the literature abounds with such measures of correlations \cite{Plenio:2007zz,Amico:2007ag,Horodecki:2009zz}. Most of them, with a notable exception being the logarithmic negativity \cite{Eisert:1998pz,Vidal:2002zz,Calabrese:2012ew,Calabrese2013}, are defined through optimization procedures, making them, at best, computationally challenging in a quantum field theory setting. 

Recently, a quantum information quantity for mixed states, dubbed reflected entropy, was introduced in \cite{Dutta:2019gen} and can be expressed simply as follows. A quantum state $\rho_{AB}$ on a bipartite Hilbert space $\H_A\otimes \H_B$ can be canonically purified%
\footnote{The archetypal example of such construction is the thermofield double state, which is the canonical purification of the thermal state.} 
as the pure state $\ket{\sqrt{\rho_{AB}}}$ in a doubled Hilbert space $(\H_A\otimes \H_B)\otimes(\H_{A^*}\otimes \H_{B^*})$. The reflected entropy $S_R(A : B)$ is then defined as the von Neumann (entanglement) entropy associated to the reduced density matrix $\rho_{AA^*} = \tr_{BB^*}(\ket{\sqrt{\rho_{AB}}}\bra{\sqrt{\rho_{AB}}})$. 
Fortunately, a replica formulation of the reflected entropy was put forward in \cite{Dutta:2019gen}, giving a practical handle for computations. This replica trick involves two replica indices, $m$ and $n$. The latter represents the usual R\'enyi index while the former generalizes the purification $\ket{\sqrt{\rho_{AB}}}$ to $\ket{\rho_{AB}^{m/2}}$, with \mbox{$m\in2\mathbb{Z}^+$}, such that%
\footnote{Note that $\ket{\rho_{AB}^{m/2}}$ is not normalized here.}
 $\tr_{A^*B^*}\big(\ket{\rho_{AB}^{m/2}}\bra{\rho_{AB}^{m/2}}\big)= \rho_{AB}^m$. 
 One then defines $\rho_{AA^*}^{(m)}$ by tracing out over $\H_{B}\otimes \H_{B^*}$ in the purified state $\ket{\rho_{AB}^{m/2}}$,
and generalizes the reflected entropy with the replica \mbox{index} $n$ in a similar manner as the R\'enyi entropies,
\be
S_R^{(n)}(A:B)=\lim_{m\rightarrow1}\frac{1}{1-n}\log\frac{\tr\big(\rho_{AA^*}^{(m)}\big)^n}{\big(\tr\rho_{AB}^m\big)^n}\,.
\ee
The (von Neumann) reflected entropy is recovered by taking the $n\rightarrow1$ limit
\be
S_R(A:B) = \lim_{n\rightarrow1} S_R^{(n)}(A:B)\,.
\ee
The reflected entropy satisfies interesting properties, some of which we list below.
\begin{itemize}[leftmargin=21pt]
\item[$(i)$] For a pure state $\rho_{AB}$, the reflected entropy reduces to twice the entanglement entropy,
\be
S_R(A:B) = 2S(A)=2S(B)\,,\quad\; \rho_{AB}\;{\rm pure}\,.
\ee
\item[$(ii)$]
For a factorized state, the reflected entropy vanishes,
\be
S_R(A:B) = 0\,,\qquad \rho_{AB}=\rho_A\otimes\rho_B\,.
\ee
\item[$(iii)$] The reflected entropy is bounded from above and below:
\be
I(A:B)\le S_R(A:B) \le 2\min\{S(A), S(B)\}\,. \lb{bounds}
\ee
\item[$(iv)$] For a tripartite pure state, the reflected entropy satisfies a polygamy inequality:  
\be
S_R(A:B) + S_R(A:C) \ge S_R(A:BC)\,. \lb{polygamy}
\ee
\end{itemize}

In the holographic context, the reflected entropy was suggested \cite{Dutta:2019gen} as a quantity that computes the (minimal) area of the entanglement wedge cross-section \cite{Takayanagi:2017knl,Nguyen:2017yqw, BabaeiVelni:2019pkw}, which can be thought of as a generalization of the Ryu-Takayanagi surface \cite{Ryu:2006bv,Ryu:2006ef}. Most of the literature available on the reflected entropy thus concerns (holographic) CFTs in two dimensions, see e.g. \cite{Jeong:2019xdr,Kusuki:2019rbk,Kusuki:2019evw,Kudler-Flam:2020url,Moosa:2020vcs}. For further developments, we refer the reader to \cite{Akers:2019gcv,Bueno:2020vnx,Asrat:2020uib,Chandrasekaran:2020qtn,Li:2020ceg}, while for candidates of multipartite reflected entropy, see \cite{Chu:2019etd,Bao:2019zqc,Marolf:2019zoo}.

The main purpose of this paper is then to study the reflected entropy in $(2+1)$-dimensional Chern-Simons field theories, and determine what topological data it encodes. We focus on mixed states that are simple to construct from a (pure) ground state, but which are still expected to reflect the essential features of generic mixed states. 
We start with a system in a pure state $\rho$, divided into three non-overlapping regions: regions $A$ and $B$, and the rest of the system, $C$. We then consider the reduced density matrix on $A\cup B$, $\rho_{AB}=\tr_C\hspace{1pt}\rho$, which is in general that of a mixed state. The entanglement structure of such mixed states of topologically ordered systems have been investigated through the lens of logarithmic negativity in, e.g., \cite{Wen:2016snr,Wen:2016bla,2013PhRvA..88d2319C,2013PhRvA..88d2318L}. The mutual information between two subsystems, being a measure of total correlations, is also a useful probe of the topological nature of systems, and was studied in \cite{Wen:2016snr} for tripartite ground states in 3$d$ Chern-Simons theories.
 
Additionally in this paper, we compute the odd entropy \cite{Tamaoka:2018ned} in $(2+1)$-dimensional Chern-Simons field theories as well.
Mainly introduced as an information-theoretic quantity that captures the entanglement wedge cross-section in two-dimensional holographic CFTs, the odd entropy has only recently been studied for itself in \cite{Kusuki:2019evw,Kudler-Flam:2020url,Angel-Ramelli:2020wfo,Mollabashi:2020ifv} for conformal and Lifshitz field theories. The reflected entropy and the odd entropy are thus related to each others in 2$d$ holographic CFTs. We compare the two quantities in 3$d$ Chern-Simons theories.

Our paper is organized as follows. In Section \ref{sec2}, we start by briefly reviewing the edge theory approach in $3d$ Chern-Simons theories. We then show how to construct a canonical purification mapping the density operator $\rho_{AB}^m$ to a purified state $\ket{\rho_{AB}^{m/2}}$ in a doubled Hilbert space $(\H_A\otimes \H_B)\otimes(\H_{A^*}\otimes \H_{B^*})$ within the edge theory framework. We subsequently compute the (R\'enyi) reflected entropy for Chern-Simons theories defined on spheres and tori. We study various bipartite mixed states obtained from tripartitions of the sphere and the torus by tracing over one of the regions. In all cases under consideration, we find that the reflected entropy agrees with the corresponding mutual information, though we note that their R\'enyi versions do not coincide in general.
We then proceed in Section \ref{sec3} to calculate the reflected entropy using surgery techniques, and find perfect agreement with the results obtained with the edge theory approach. Section \ref{sec4} presents our results on the odd entropy and its `regulated' form. Our main motivation for considering the latter comes from its holographic dual as the entanglement wedge cross-section, similar to the reflected entropy. 
We discuss our results in Section \ref{conclu}, and give an outlook on future research directions. Two appendices complete this work: Appendix \ref{Apdx1} contains details about the mutual information, while Appendix \ref{Apdx2} displays figures related to the calculation of reflected entropy using the surgery method.

\section{Reflected entropy via the edge theory approach} \lb{sec2}

Within the bulk-edge correspondence \cite{Witten:1988hf,Elitzur:1989nr,Moore:1991ks,PhysRevLett.101.010504,PhysRevLett.108.196402,PhysRevB.88.245137,Cano:2014pya} in topological quantum fields theories (TQFTs), boundary states in ($1+1$)--dimensional CFTs can be used to describe the reduced density matrices of ($2+1$)--dimensional topologically ordered phases. This duality can be understood from the equivalence of the modular Hamiltonian of the bulk theory with the Hamiltonian of the chiral CFT living on the boundary (e.g.\,\,the entangling surface). Consider a topological state on the 2-sphere with an entangling cut along the equator. In the `cut-and-glue' picture of \cite{PhysRevLett.108.196402}, one treats the entangling surface as a physical cut, which splits the sphere into two hemispheres $A$ (left) and $B$ (right) that possess edge states of opposite chirality propagating at their boundaries. Now, turning on a small enough (RG-relevant) coupling between the two gapless edge modes will gap out and heal the cut without affecting the gapped bulk states. One can then show that the entanglement properties between the subsystems $A$ and $B$ is reduced to those between the left and right moving edge modes. Tracing out the degrees of freedom in, e.g., subsystem $B$, therefore amounts to tracing out the right moving modes.

This cut-and-glue procedure can be interpreted \cite{PhysRevLett.108.196402} as a sudden quantum quench scenario which can be solved \cite{PhysRevLett.96.136801,Calabrese:2007rg} applying boundary CFT techniques \cite{Cardy:1989ir,Cardy:2004hm}.
The ground state of a $(1+1)$--dimensional CFT describing the coupled edges may then be obtained in terms of conformal boundary states. 
These conformally invariant boundary states are generically linear combinations of Ishibashi states $\vert h_a\dra$, and are non-normalizable. A way to regularize their norm is to perform an Euclidean time evolution by $e^{-\eps H}$, where $\eps$ is interpreted as a UV cutoff.
We will thus work with the following regularized boundary states \cite{Wen:2016snr}:
\be
\ket{\B} = \sum_a \psi_a \vert\h_a\dra\,,\qquad \vert \h_a\dra:=\frac{e^{-\eps H}}{\sqrt{\n_a}}\vert h_a\dra\,, \quad\lb{reg}
\ee
where $\psi_a$ is a complex number which depends on the choice of ground state of the Chern-Simons field theory, and $\n_a$ is a normalization factor such that \mbox{$\dla \h_a\vert \h_b\dra=\delta_{ab}$}. The Ishibashi states $\vert h_a\dra$ are the solution to the conformal boundary condition $L_n\ket{b}=\bar{L}_{-n}\ket{b}$, where $L_n$ is the generator of chiral conformal transformations, and they can be expressed in terms of the orthonormal bases $\ket{h_a,N}$ and $\ket{\overline{h_a,N}}$, usually referred to as left and right bases respectively,
\be
\vert h_a\dra = \sum_N\ket{h_a,N}\otimes\ket{\overline{h_a,N}}\,.\quad
\ee
Here $a$ labels the primary sector with conformal weight $h_a$ (corresponding to the type of quasiparticle in the TQFT), and the sum $N$ is over descendants.
 The Hamiltonian is taken to be
\be
H=\frac{2\pi}{\ell}\Big(L_0+\bar{L}_0-\frac{c}{12}\Big)\,, \lb{H}
\ee
where $\ell$ is the length of the circle on which the state $\ket{\B}$ is defined, e.g., the entangling surface between two spatial regions, and $c$ is the central charge of the underlying CFT. 
Using the fact that $L_0\ket{h_a,N}=(h_a+N)\ket{h_a,N}$ and $h_a=\bar{h}_{\bar{a}}$, the normalization factor $\n_a$ is found to be
\be
\n_a =  \chi_{h_a}\big( e^{-\frac{8\pi \eps}{\ell}}\big)\,,\lb{na}
\ee
where $\chi_{h_a}$ are the characters of the highest weight representations of the primaries $h_a$. 
The modular transformation property of the character $\chi$ in CFT reads
\be
\chi_{h_a}\big( e^{-\frac{8\pi \eps}{\ell}}\big) = \sum_b \mathcal{S}_{ab}\hspace{1pt}\chi_{h_b}\big( e^{-\frac{\pi \ell}{2\eps}}\big)\,,\lb{modtrans}
\ee
with $\mathcal{S}_{ab}$ being the matrix elements of the modular $\S$ matrix. In the thermodynamic limit $\ell/\eps\rightarrow\infty$, using \eqref{modtrans}, one finds that only the identity field (labeled by “0”) survives,
\be
\lim_{\ell/\eps\rightarrow\infty}\chi_{h_a}\big( e^{-\frac{8\pi \eps}{\ell}}\big) \simeq e^{\frac{\pi c \ell}{48\eps}}\S_{a0}\,.
\ee

\medskip
\noindent\textBF{Left-right entanglement entropy}

Before discussing how to compute the reflected entropy using the edge theory approach, we reproduce here the calculation of the left-right entanglement entropy \cite{Das:2015oha} for the regularized state \eqref{reg}, as done in \cite{Wen:2016snr}. This corresponds, for example, to the geometry in Fig.\,\hyperref[fig1]{\ref{fig1}(a)}. 

The reduced density matrix associated to the left-moving sector is
\be
\rho_L = \tr_R\left(\ket{\B}\bra{\B}\right)=: \sum_a \vert\psi_a\vert^2 \rho_{L,a}\,, \lb{rhoL}
\ee
where we defined
\be
\rho_{L,a}&=&\frac{1}{\n_a}\sum_{N}e^{-\frac{8\pi \eps}{\ell}(h_a+N-\frac{c}{24})}\ket{h_a,N}\bra{h_a,N}\,.\qquad
\ee
The $n^{\rm th}$ power of the reduced density matrix $\rho_L$ reads
\be
\rho_L^n=\sum_a \vert\psi_a\vert^{2n} \rho_{L,a}^n\,,
\lb{rhoEEn}
\ee
with
\be
\rho_{L,a}^n = \frac{1}{\n_a^n}\sum_{N}e^{-\frac{8\pi n \eps}{\ell}(h_a+N-\frac{c}{24})}\ket{h_a,N}\bra{h_a,N}\,.\qquad
\ee
Taking the trace of \eqref{rhoEEn}, we end up with
\be
\Tr\rho_L^n&=&\sum_a \vert\psi_a\vert^{2n} \Tr\rho_{L,a}^n\,, \nn\\
&=& \sum_a \vert\psi_a\vert^{2n}\frac{\chi_{h_a}\big( e^{-\frac{8\pi n \eps}{\ell}}\big)}{\left(\chi_{h_a}\big( e^{-\frac{8\pi \eps}{\ell}}\big) \right)^n}\,,
\ee
where we have used \eqref{na}. In the thermodynamic limit $\ell/\eps\rightarrow\infty$ one finds
\be
\Tr\rho_L^n \simeq e^{\frac{\pi c \ell}{48\eps}(\frac{1}{n}-n)}\sum_a \vert\psi_a\vert^{2n} (\S_{a0})^{1-n}\,.
\lb{trrhoEE}
\ee
The R\'enyi entropies thus read
\be
S^{(n)}(L) &=& \frac{1}{1-n}\log\frac{\Tr\rho_L^n}{(\Tr\rho_L)^n}\,,  \nn\\
&=&\Big(1+\frac{1}{n}\Big) \frac{\pi c}{48}\frac{\ell}\eps + \frac{1}{1-n}\log\frac{\sum_a \vert\psi_a\vert^{2n} (\S_{a0})^{1-n}}{\big(\sum_a \vert\psi_a\vert^{2}\big)^n},\nn\\ \lb{LRRE}
\ee
and the von Neumann entropy $S(L) = \lim_{n\rightarrow1} S^{(n)}(L)$ is
\be
S(L) &=& \frac{\pi c}{24}\frac{\ell}\eps +\frac{\sum_a \vert\psi_a\vert^{2} \log\S_{a0}}{\sum_a \vert\psi_a\vert^{2}} -\frac{\sum_a \vert\psi_a\vert^{2} \log\vert\psi_a\vert^{2}}{\sum_a \vert\psi_a\vert^{2}}\qquad\nonumber\\
&& \hspace{0.75cm}  +\log \sum_a \vert\psi_a\vert^{2}\,.
\ee
Throughout this paper, we will work with normalized states such that $\sum_a \vert\psi_a\vert^{2}=1$, and the above expressions can be simplified to 
\begin{align}
S^{(n)}(L)&=\Big(1+\frac{1}{n}\Big) \frac{\pi c}{48}\frac{\ell}\eps + \frac{1}{1-n}\log\sum_a \vert\psi_a\vert^{2n} (\S_{a0})^{1-n},\quad\;\;\,\nn\\
S(L) &= \frac{\pi c}{24}\frac{\ell}\eps +\sum_a \vert\psi_a\vert^{2} \log\S_{a0} -\sum_a \vert\psi_a\vert^{2} \log\vert\psi_a\vert^{2}\,.\nn\\[-.5\baselineskip]\lb{LREE}
\end{align}
The first term in \eqref{LREE} obeys the area law, while the third piece takes the form of the Shannon entropy of the coefficients of the choice of state. The last two terms constitute the celebrated topological entanglement entropy \cite{Kitaev:2005dm,Levin:2006zz}, which is finite and universal, and they depend on the topology of the system as well as the choice of ground state. It may also be expressed in terms of the quantum dimensions $d_a= \S_{a0}/S_{00}$, and the total quantum dimension $\mathcal{D}=1/\S_{00}=(\sum_a d_a^2)^{1/2}$.

\medskip
\noindent\textBF{Left-right reflected entropy (bipartite pure state)}

To compute the reflected entropy between the chiral and anti-chiral edge modes using the replica trick, we must obtain the purification $\ket{\rho^{m/2}}$ of $\rho^m$ in a doubled Hilbert space $(\H_L\otimes \H_R)\otimes(\H_{L^*}\otimes \H_{R^*})$. Note that here $\rho=\ket{\B}\bra{\B}$ is already pure, but the construction of this `purification' is a necessary exercise for later purpose. The first step is to compute $\rho^{m/2}$, with $m\in 2\mathbb{Z}^+$, which is simply $\rho^{m/2}=\rho$ since $\rho$ is idempotent, i.e.
\begin{align}
\hspace{-6pt}\rho^{m/2} &= \sum_{a,a'}\psi_a\psi_{a'}^* \vert \h_a\dra\dla\h_{a'}\vert\,,\quad\nn\\
&= \sum_{a,a'}\hspace{-2pt} \frac{\psi_a\psi_{a'}^*}{\sqrt{\n_a\n_{a'}}}\hspace{-2pt}\sum_{N}\hspace{-2pt}\sum_{N'}\hspace{-2.pt}e^{-\frac{4\pi \eps}{\ell}(h_a+N-\frac{c}{24})-\frac{4\pi \eps}{\ell}(h_{a'}+N'-\frac{c}{24})}\nn\\
&\hspace{1.5cm} \times\ket{h_a,N}\ket{\overline{h_a,N}}\bra{h_{a'},N'}\bra{\overline{h_{a'},N'}}\,.
\end{align}
A canonical doubling of the Hilbert space provides the simplest purification $\ket{\rho^{m/2}}$ on $\H_L\otimes \H_R\otimes\H_{L^*}\otimes \H_{R^*}$ as follows
\begin{align}
\ket{\rho^{m/2}}&:= \nn\\
&\hspace{-15pt}\sum_{a,a'} \frac{\psi_a\psi_{a'}}{\sqrt{\n_a\n_{a'}}}\sum_{N}\sum_{N'}e^{-\frac{4\pi \eps}{\ell}(h_a+N-\frac{c}{24})-\frac{4\pi \eps}{\ell}(h_{a'}+N'-\frac{c}{24})}\nn\\
&\hspace{0.7cm}\times \underbrace{\ket{h_a,N}\ket{\overline{h_a,N}}}_{\in\, \H_L\otimes \H_R}\otimes\underbrace{\ket{h_{a'},N'}\ket{\overline{h_{a'},N'}}}_{\in\, \H_{L^*}\otimes \H_{R^*}}\,.\nn\qquad\quad\\[-.8\baselineskip]&\,
\end{align}
It is then straightforward to compute the reduced density matrix $\rho_{LL^*}^{(m)}$, i.e.
\be
\rho_{LL^*}^{(m)} = \rho_L\otimes\rho_{L^*}\,,
\ee
hence
\be
\tr\big(\rho_{LL^*}^{(m)}\big)^n=  (\Tr\rho_L^n)^2\,,
\ee
where $\rho_L$ is defined in \eqref{rhoL} and $\Tr\rho_L^n$ is given by \eqref{trrhoEE}. The R\'enyi reflected entropy is thus given by twice the left-right R\'enyi entropy,
\be
S_R^{(n)}(L:R) = 2S^{(n)}(L)\,,\quad \lb{pureS}
\ee
 as expected for pure states.

\subsection{Sphere}

\begin{figure}[t]
\centering
\includegraphics[scale=0.9]{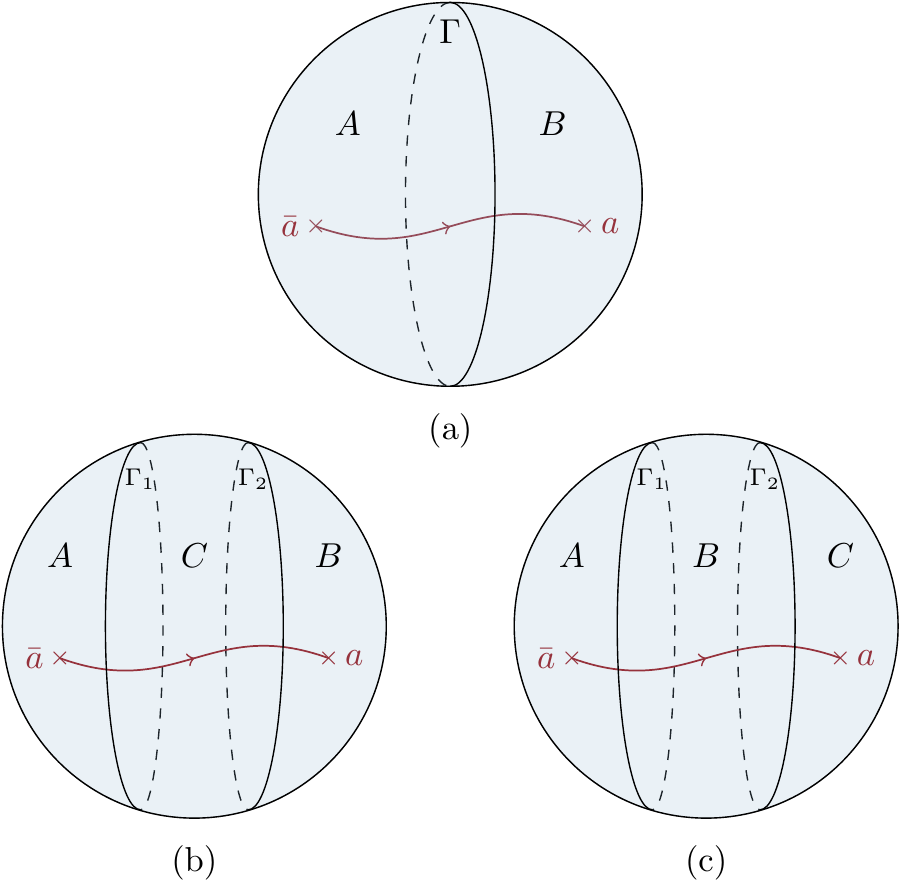}
\caption{Different states prepared on the 2-sphere. In each of them, a Wilson line connecting two conjugate quasiparticles threads through all interfaces $\Gamma_i$ separating the subsystems. (a) Bipartite system with one interface $\Gamma$ separating $A$ from $B$. (b) Tripartite system with $A$ disconnected from $B$ by $C$. \mbox{(c) Tripartite} system with adjacent $A$ and $B$.}
   \label{fig1}
\end{figure}

We consider here a Chern-Simons theory which lives on the 2-sphere. We are interested in the reflected entropy between the subsystems $A$ and $B$, as for example depicted in Fig.\,\ref{fig1}. We assume that there are two quasiparticles on the sphere, i.e.\,\,one Wilson line threading through the interfaces $\Gamma_{i}$. In these cases, the boundary state can be expressed as 
\be
\ket\B = \sum_a\psi_a\bigotimes_{i=1}^M\vert\h_a^{i}\dra\,,
\ee
with $i$ labeling the $M$ interfaces $\Gamma_i$ of length $\ell_i$ between the different subsystems, and 
\be
&&\vert \h_a^i\dra= \frac{e^{-\eps H_i}}{\sqrt{\n_a^{i}}}\vert h_a^i\dra\,,\\
&&H_i\,=\frac{2\pi}{\ell_i}\Big(L_0^i+\bar{L}_0^i-\frac{c}{12}\Big)\,,\\
&&\n_a^{i} =  \chi_{h_a}\big( e^{-\frac{8\pi \eps}{\ell_i}}\big)\,.
\ee
Note that the vacuum state (i.e.\;no Wilson line) corresponds to setting $\psi_a=\delta_{a0}$.

\subsubsection{Two disjoint regions}

Let us first focus on the situation represented in Fig.\,\hyperref[fig1]{\ref{fig1}(b)}, where $A$ and $B$ are separated by a third subsystem $C$. There are two entangling cuts $\Gamma_{1}$ and $\Gamma_{2}$ separating $A$ and $C$ and $C$ and $B$, respectively.  There is at most one Wilson line that threads through both interfaces.
The boundary state may be expressed as \be
\ket\B = \sum_a\psi_a\vert\h_a^{1}\dra\otimes\vert\h_a^2\dra\,, \lb{Bdisj}
\ee
from which one obtains the reduced density matrix $\rho_{AB}$ by tracing over the modes in $C$, that is
\be
\rho_{AB}&=&\sum_a\vert\psi_a\vert^2 \rho_{A,a}\otimes\rho_{B,a}\,, \lb{rhoSdisj}
\ee
where
\begin{align}
\rho_{A,a} &=\frac{1}{\n_a^1}\sum_{N_1}e^{-\frac{8\pi \eps}{\ell_1}(h_a+N_1-\frac{c}{24})}\ket{h_a,N_1}\bra{h_a,N_1}\,,\nn\\
\rho_{B,a} &=\frac{1}{\n_a^2}\sum_{N_2}e^{-\frac{8\pi \eps}{\ell_2}(h_a+N_2-\frac{c}{24})}\ket{\overline{h_a,N_2}}\bra{\overline{h_a,N_2}}\,.\nn\\[-.6\baselineskip]
\end{align}
Next we compute $\rho_{AB}^{m/2}$ for even positive $m$, 
\be
\rho_{AB}^{m/2}&=&\sum_a\vert\psi_a\vert^{m} \rho_{A,a}^{m/2}\otimes\rho_{B,a}^{m/2}\,,
\ee
where
\begin{align}
\rho_{A,a}^{m/2} &=\frac{1}{(\n_a^1)^{m/2}}\sum_{N_1}e^{-\frac{4m\pi \eps}{\ell_1}(h_a+N_1-\frac{c}{24})}\ket{h_a,N_1}\bra{h_a,N_1}\,,\nn\\
\rho_{B,a}^{m/2} &=\frac{1}{(\n_a^2)^{m/2}}\sum_{N_2}e^{-\frac{4m\pi \eps}{\ell_2}(h_a+N_2-\frac{c}{24})}\ket{\overline{h_a,N_2}}\bra{\overline{h_a,N_2}}\,.\;\nn\\[-.6\baselineskip]
\end{align}
Then, to construct the purification $\ket{\rho_{AB}^{m/2}}$, we turn the bras in $\rho_{A,a}^{m/2}$ and $\rho_{B,a}^{m/2}$ into kets in $\H_{A^*}$ and $\H_{B^*}$, respectively. The purified state thus reads
\be
\ket{\rho_{AB}^{m/2}} = \sum_a\vert\psi_a\vert^{m} \ket{\rho_{A,a}^{m/2}}\otimes\ket{\rho_{B,a}^{m/2}}\,,
\ee
with
\begin{align}
\ket{\rho_{A,a}^{m/2}} &=\frac{1}{(\n_a^1)^{m/2}}\sum_{N_1}e^{-\frac{4m\pi \eps}{\ell_1}(h_a+N_1-\frac{c}{24})}\nn\\
&\hspace{3cm} \times\underbrace{\ket{h_a,N_1}}_{\in\, \H_A}\otimes\underbrace{\ket{h_a,N_1}}_{\in\,\H_{A^*}}\,,\nn\\
\ket{\rho_{B,a}^{m/2}} &=\frac{1}{(\n_a^2)^{m/2}}\sum_{N_2}e^{-\frac{4m\pi \eps}{\ell_2}(h_a+N_2-\frac{c}{24})}\nn\\
&\hspace{3cm} \times\underbrace{\ket{\overline{h_a,N_2}}}_{\in\, \H_B}\otimes\underbrace{\ket{\overline{h_a,N_2}}}_{\in\,\H_{B^*}}\,.\;\;
\end{align}
The reduced density matrix for $AA^*$ may then be written as
\be \lb{pure_ensAA}
\rho_{AA^*}^{(m)}= \sum_a\vert\psi_a\vert^{2m} \frac{\chi_{h_a}\big( e^{-\frac{8\pi m \eps}{\ell_2}}\big)}{\big(\chi_{h_a}\big( e^{-\frac{8\pi \eps}{\ell_2}}\big)\big)^m}\ket{\rho_{A,a}^{m/2}}\bra{\rho_{A,a}^{m/2}}\,,\qquad
\ee
from which follows 
\begin{align}
\tr\big(\rho_{AA^*}^{(m)}\big)^n &= \sum_a\vert\psi_a\vert^{2nm} \prod_{i=1,2}\frac{\big(\chi_{h_a}\big( e^{-\frac{8\pi m \eps}{\ell_i}}\big)\big)^n}{\big(\chi_{h_a}\big( e^{-\frac{8\pi \eps}{\ell_i}}\big)\big)^{nm}}\,,\nn\\
&\simeq e^{\frac{\pi c}{48}\frac{\ell_1+\ell_2}{\eps}(\frac{n}{m}-nm)}\sum_a\vert\psi_a\vert^{2nm} (\S_{a0})^{2n(1-m)}\,,\;\nn\\[-.6\baselineskip]\lb{rhoSdis}
\end{align}
where we took the thermodynamic limit $\ell_i/\eps\rightarrow\infty$ in the second line.
Finally, we obtain the R\'enyi and von Neumann reflected entropies 
\be
S_R^{(n)}(A:B) &=& \frac{1}{1-n}\log\sum_a\vert\psi_a\vert^{2n}\,,\\
S_R(A:B) &=& -\sum_a \vert\psi_a\vert^{2} \log\vert\psi_a\vert^{2}\,,\quad\lb{disjointSRS}\\
&=& I(A:B)\,.\nn
\ee
The area-law terms disappear in both $S_R^{(n)}$ and $S_R$, while their universal parts result only from the fluctuations of the Wilson line, given in $S_R$ by the Shannon entropy of the classical probability distribution. Note that for density matrices which are the mixtures of factorized states, as in \eqref{rhoSdisj}, the corresponding (R\'enyi) reflected entropy is alway given as above. Interestingly, we observe that the reflected entropy coincides with the mutual information, see \eqref{MISD} in Appendix \ref{Apdx1}. Note that formally $S_R^{(n)}\neq I^{(n)}$ for $n>1$,  though for the Abelian Chern-Simons theories the two quantities are equal.

\subsubsection{Two adjacent regions}

For the case of adjacent $A$ and $B$, as shown in Fig.\,\hyperref[fig1]{\ref{fig1}(c)}, the ground state is again \eqref{Bdisj}. The reduced density matrix for the subsystem $A\cup B$ reads
\be
\rho_{AB}&=&\sum_a\vert\psi_a\vert^2 \rho_{AB,a}^{\Gamma_1}\otimes\rho_{B,a}^{\Gamma_2}\,, \lb{rhoSadj}
\ee
where
\begin{align}
&\rho_{AB,a}^{\Gamma_1} \notag\\
&\hspace{0.5cm}=\frac{1}{\n_a^1}\sum_{N_1}\sum_{N_1'}e^{-\frac{4\pi \eps}{\ell_1}(h_a+N_1-\frac{c}{24})}e^{-\frac{4\pi \eps}{\ell_1}(h_a+N_1'-\frac{c}{24})}\notag\\
&\hspace{2cm}\times\ket{h_a,N_1}\ket{\overline{h_a,N_1}}\bra{h_a,N_1'}\bra{\overline{h_a,N_1'}}\,,\notag\\
&\rho_{B,a}^{\Gamma_2} =\frac{1}{\n_a^2}\sum_{N_2}\hspace{-3pt}e^{-\frac{8\pi \eps}{\ell_2}(h_a+N_2-\frac{c}{24})}\ket{\overline{h_a,N_2}}\bra{\overline{h_a,N_2}}\,.\notag\\[-.6\baselineskip]&\,
\end{align}
As we did in the previous case, to construct the canonical purification $\ket{\rho_{AB}^{m/2}}$ of $\rho_{AB}^m$, we first compute $\rho_{AB}^{m/2}$ and then flip the bras to kets for basis in $\H_{A^*}\otimes\H_{B^*}$. We then find 
\be
\ket{\rho_{AB}^{m/2}}&=&\sum_a\vert\psi_a\vert^{m} \ket{\rho_{AB,a}^{\Gamma_1}}\otimes\ket{(\rho_{B,a}^{\Gamma_2})^{m/2}}\,,\quad
\ee
where
\begin{align}
&\ket{\rho_{AB,a}^{\Gamma_1}}=\frac{1}{\n_a^1}\sum_{N_1}\sum_{N_1'}e^{-\frac{4\pi \eps}{\ell_1}(h_a+N_1-\frac{c}{24})}e^{-\frac{4\pi \eps}{\ell_1}(h_a+N_1'-\frac{c}{24})}\notag\\
&\hspace{2.5cm}\times\underbrace{\ket{h_a,N_1}\ket{\overline{h_a,N_1}}}_{\in\, \H_A\otimes \H_B}\otimes\underbrace{\ket{h_a,N_1'}\ket{\overline{h_a,N_1'}}}_{\in\, \H_{A^*}\otimes \H_{B^*}},\notag\\
&\ket{(\rho_{B,a}^{\Gamma_2})^{m/2}} =\frac{1}{(\n_a^2)^{m/2}}\sum_{N_2}\hspace{-3pt}e^{-\frac{4m\pi \eps}{\ell_2}(h_a+N_2-\frac{c}{24})}\notag\\
&\hspace{4cm}\times\underbrace{\ket{\overline{h_a,N_2}}}_{\in\, \H_B}\otimes\underbrace{\ket{\overline{h_a,N_2}}}_{\in\,\H_{B^*}}\,.\quad\notag\\[-.6\baselineskip]&\,
\end{align}
The reduced density matrix for $AA^*$ can be easily calculated,
\be
\rho_{AA^*}^{(m)}= \sum_a\vert\psi_a\vert^{2m} \frac{\chi_{h_a}\big( e^{-\frac{8\pi m \eps}{\ell_2}}\big)}{\big(\chi_{h_a}\big( e^{-\frac{8\pi \eps}{\ell_2}}\big)\big)^m} \rho_{AA^*,a}^{\Gamma_1}\,,\quad
\ee
where
\begin{align}
&\rho_{AA^*,a}^{\Gamma_1}=\frac{1}{(\n_a^1)^2}\sum_{N}\sum_{N'}e^{-\frac{8\pi \eps}{\ell_1}(h_a+N-\frac{c}{24})}e^{-\frac{8\pi \eps}{\ell_1}(h_a+N'-\frac{c}{24})}\notag\\
&\hspace{2.4cm}\times\ket{h_a,N}\otimes\ket{h_a,N'}\bra{h_a,N}\otimes\bra{h_a,N'}\,,\quad\nn\\[.1\baselineskip]
\end{align}
from which follows 
\begin{align}
\tr\big(\rho_{AA^*}^{(m)}\big)^n &\nonumber\\
&\hspace{-1.25cm}= \sum_a\vert\psi_a\vert^{2nm} \frac{\big(\chi_{h_a}\big( e^{-\frac{8\pi n \eps}{\ell_1}}\big)\big)^2}{\big(\chi_{h_a}\big( e^{-\frac{8\pi \eps}{\ell_1}}\big)\big)^{2n}}\frac{\big(\chi_{h_a}\big( e^{-\frac{8\pi m \eps}{\ell_2}}\big)\big)^n}{\big(\chi_{h_a}\big( e^{-\frac{8\pi \eps}{\ell_2}}\big)\big)^{nm}}\,,\nonumber \\
&\hspace{-1.25cm}\simeq e^{\frac{\pi c\ell_1}{24\eps}(\frac{1}{n}-n)}e^{\frac{\pi c\ell_2}{48\eps}(\frac{n}{m}-nm)}\sum_a\vert\psi_a\vert^{2nm} (\S_{a0})^{2-n(1+m)}\,,\nn\\[-.6\baselineskip]
\end{align}
where we took the thermodynamic limit $\ell_i/\eps\rightarrow\infty$ in the second line. Finally, we obtain the R\'enyi and von Neumann reflected entropies 
\begin{align}
&S_R^{(n)}(A:B) \notag\\
&\hspace{0.5cm} =\Big(1+\frac{1}{n}\Big) \frac{\pi c}{24}\frac{\ell_1}\eps + \frac{1}{1-n}\log\sum_a \vert\psi_a\vert^{2n} (\S_{a0})^{2(1-n)}\,,\notag\\[-.6\baselineskip]\\ \lb{sphere-adjacent}
&S_R(A:B) \notag\\ 
&\hspace{0.5cm}= \frac{\pi c}{12}\frac{\ell_1}\eps +2\sum_a \vert\psi_a\vert^{2} \log\S_{a0} -\sum_a \vert\psi_a\vert^{2} \log\vert\psi_a\vert^{2}\,.\qquad\nonumber\\[-.6\baselineskip]
\end{align}
We notice that the area-law terms do not cancel in this case. Furthermore, we find that
\be
S_R^{(n)}(A:B) = I^{(n)}(A:B)\,.
\ee
Not only the reflected entropy matches the mutual information, but their R\'enyi generalizations do also.

\subsection{Torus}
\begin{figure}[b]
\centering
\includegraphics[scale=0.78]{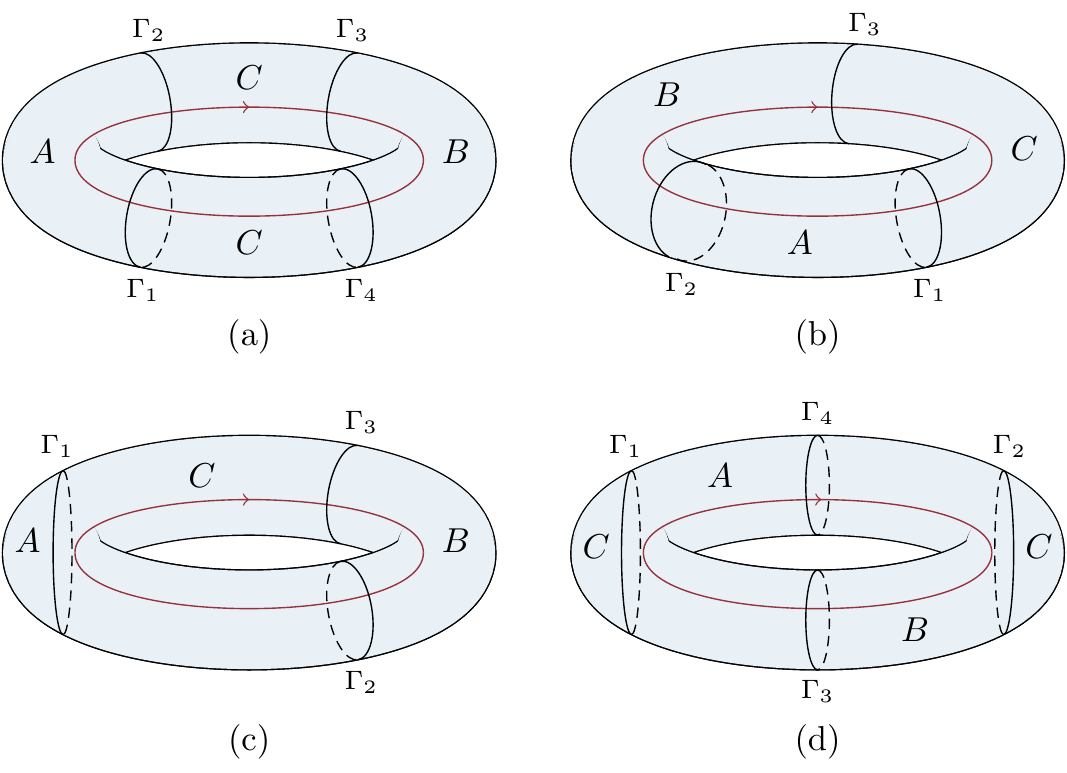}
\caption{Different tripartite states on the 2-torus. Two regions $A$ and $B$ with complementary non-contractible [(a)--(c)] or contractible (d) region $C$, and with a longitudinal \mbox{Wilson} loop tunneling through non-contractible regions only. (a) Two disjoint non-contractible regions, (b) Two adjacent non-contractible regions with a one component interface, (c) Two disjoint regions with contractible $A$ and non-contractible $B$. (d) Two adjacent non-contractible regions with a two component interface, with contractible $C$.}
   \label{fig2}
\end{figure}

We consider now a tripartite pure state of a Chern-Simons theory which lives on the 2-torus. We are interested in the reflected entropy between the non-complementary subsystems $A$ and $B$, as for example depicted in Fig.\,\ref{fig2}. The third subsystem is denoted by $C$. A Wilson loop threads through the interfaces $\Gamma_{i}$, and can in general fluctuate among different topological sectors $a$ with probability $\vert\psi_a\vert^2$. 

\subsubsection{Two disjoint non-contractible regions}
Let us first focus on the situation represented in Fig.\,\hyperref[fig2]{\ref{fig2}(a)}, where $A$ and $B$ are separated by a third subsystem $C$. There are four entangling cuts $\Gamma_{1,\cdots,4}$ separating $A$ and $B$ from $C$.  A Wilson loop pierces through all the interfaces.
The boundary state may be expressed as \be
\ket\B = \sum_a\psi_a\bigotimes_{i=1}^4\vert\h_a^{i}\dra\,.
\ee
The reduced density matrix for the subsystem $A\cup B$ is easily found to be
\be
\rho_{AB}&=&\sum_a\vert\psi_a\vert^2 \rho_{A,a}^{\Gamma_1}\otimes\rho_{A,a}^{\Gamma_2}\otimes\rho_{B,a}^{\Gamma_3}\otimes\rho_{B,a}^{\Gamma_4}\,,
\ee
where
\begin{align}
\rho_{A,a}^{\Gamma_1} &=\frac{1}{\n_a^1}\sum_{N_1}e^{-\frac{8\pi \eps}{\ell_1}(h_a+N_1-\frac{c}{24})}\ket{h_a,N_1}\bra{h_a,N_1}\,,\nn\\
\rho_{A,a}^{\Gamma_2} &=\frac{1}{\n_a^2}\sum_{N_2}e^{-\frac{8\pi \eps}{\ell_2}(h_a+N_2-\frac{c}{24})}\ket{\overline{h_a,N_2}}\bra{\overline{h_a,N_2}}\,,\nn\\
\rho_{B,a}^{\Gamma_3} &=\frac{1}{\n_a^3}\sum_{N_3}e^{-\frac{8\pi \eps}{\ell_3}(h_a+N_3-\frac{c}{24})}\ket{h_a,N_3}\bra{h_a,N_3}\,,\nn\\
\rho_{B,a}^{\Gamma_4} &=\frac{1}{\n_a^4}\sum_{N_4}e^{-\frac{8\pi \eps}{\ell_4}(h_a+N_4-\frac{c}{24})}\ket{\overline{h_a,N_4}}\bra{\overline{h_a,N_4}}\,.\nn\\[-.6\baselineskip]
\end{align}
Since $\rho_{AB}$ is a classically correlated mixed state, the reduced density matrix for the subsystem $AA^*$ is an ensemble of pure states (similarly as in \eqref{pure_ensAA}),
and the R\'enyi and von Neumann reflected entropies thus read
\be
S_R^{(n)}(A:B) &=& \frac{1}{1-n}\log\sum_a\vert\psi_a\vert^{2n}\,,\lb{SRTdis}\\
S_R(A:B) &=&  -\sum_a \vert\psi_a\vert^{2} \log\vert\psi_a\vert^{2}\,,\quad\\
&=& I(A:B)\,.\nn
\ee
As for the sphere, in the torus case the reflected entropy equals the mutual information which was computed in \cite{Wen:2016snr} (see also \eqref{MITG} in Appendix \ref{Apdx1}), and only retains the Shannon entropy arising from the classical distribution $\{\vert\psi_a\vert^2\}$.

\subsubsection{Two adjacent non-contractible regions}
For the geometry depicted in Fig.\,\hyperref[fig2]{\ref{fig2}(b)}, $A$ and $B$ are adjacent with a one-component interface between them. There are three entangling cuts $\Gamma_{1}$, $\Gamma_{2}$, and $\Gamma_{3}$ separating $A$ and $C$, $B$ and $A$, and $C$ and $B$, respectively, with a Wilson loop
threading through them all. The boundary state may be expressed as \be
\ket\B = \sum_a\psi_a\bigotimes_{i=1}^3\vert\h_a^{i}\dra\,.
\ee
The reduced density matrix $\rho_{AB}$ reads
\be
\rho_{AB}&=&\sum_a\vert\psi_a\vert^2 \rho_{A,a}^{\Gamma_1}\otimes\rho_{AB,a}^{\Gamma_2}\otimes\rho_{B,a}^{\Gamma_3}\,,
\ee
where
\begin{align}
&\rho_{AB,a}^{\Gamma_2}=\frac{1}{\n_a^2}\sum_{N_2}\sum_{N_2'}e^{-\frac{4\pi \eps}{\ell_2}(h_a+N_2-\frac{c}{24})}e^{-\frac{4\pi \eps}{\ell_2}(h_a+N_2'-\frac{c}{24})}\notag\\
&\hspace{2.35cm}\times\ket{h_a,N_2}\ket{\overline{h_a,N_2}}\bra{h_a,N_2'}\bra{\overline{h_a,N_2'}}\,,\notag\\
&\rho_{A,a}^{\Gamma_1} =\frac{1}{\n_a^1}\sum_{N_1}e^{-\frac{8\pi \eps}{\ell_1}(h_a+N_1-\frac{c}{24})}\ket{h_a,N_1}\bra{h_a,N_1}\,,\notag\\
&\rho_{B,a}^{\Gamma_3} =\frac{1}{\n_a^3}\sum_{N_3}e^{-\frac{8\pi \eps}{\ell_3}(h_a+N_3-\frac{c}{24})}\ket{\overline{h_a,N_3}}\bra{\overline{h_a,N_3}}\,.\notag\\[-.8\baselineskip]&\,
\end{align}
Following the procedures discussed in the previous sections, one can construct the canonical purification $\ket{\rho_{AB}^{m/2}}$ and compute its associated reduced density matrix $\rho_{AA^*}^{(m)}$.
Then one obtains, by taking the thermodynamic limit,
\begin{align}
\tr\big(\rho_{AA^*}^{(m)}\big)^n &\nonumber\\
&\hspace{-1.6cm}= \sum_a\vert\psi_a\vert^{2nm} \frac{\big(\chi_{h_a}\big( e^{-\frac{8\pi n \eps}{\ell_2}}\big)\big)^2}{\big(\chi_{h_a}\big( e^{-\frac{8\pi \eps}{\ell_2}}\big)\big)^{2n}}\prod_{i=1,3}\frac{\big(\chi_{h_a}\big( e^{-\frac{8\pi m \eps}{\ell_i}}\big)\big)^n}{\big(\chi_{h_a}\big( e^{-\frac{8\pi \eps}{\ell_i}}\big)\big)^{nm}}\,,\nonumber\\
&\hspace{-1.6cm}\simeq e^{\frac{\pi c \ell_2}{24\eps}(\frac{1}{n}-n)}e^{\frac{\pi c(\ell_1+\ell_3)}{48\eps}(\frac{n}{m}-nm)}\sum_a\vert\psi_a\vert^{2nm} (\S_{a0})^{2(1-nm)}.\nn\\[-.6\baselineskip]
\end{align}
The R\'enyi and von Neumann reflected entropies are expressed as
\begin{align}
&S_R^{(n)}(A:B) \notag\\
&\hspace{0.5cm} =\Big(1+\frac{1}{n}\Big) \frac{\pi c}{24}\frac{\ell_2}\eps + \frac{1}{1-n}\log\sum_a \vert\psi_a\vert^{2n} (\S_{a0})^{2(1-n)}\,,\notag\\[-.6\baselineskip]\lb{SRRadj}\\
&S_R(A:B) \notag\\ 
&\hspace{0.5cm}= \frac{\pi c}{12}\frac{\ell_2}\eps +2\sum_a \vert\psi_a\vert^{2} \log\S_{a0}-\sum_a \vert\psi_a\vert^{2} \log\vert\psi_a\vert^{2}\,.\nn\\[-.6\baselineskip]\lb{SRadj}
\end{align}
The (R\'enyi) mutual information for the adjacent configuration has been computed in \cite{Wen:2016snr} (see also \eqref{MITG} in  Appendix \ref{Apdx1}). We find that 
\be
S_R^{(n)}(A:B) = I^{(n)}(A:B)\,,
\ee
as on the 2-sphere.

\subsubsection{Non-contractible multi-component interfaces}

We can now consider the more general case where $A$ and $B$ are each composed of an arbitrary number of non-contractible components with an arbitrary number of shared interfaces between them. It is convenient to think in terms of these interfaces. 
There are $M$ interfaces $\Gamma_i$ in total, of three types: $M_{AB}$ between $A$ and $B$, $M_A$ between $A$ and $C$, and $M_B$ between $B$ and $C$, where $C$ is the complementary non-contractible subsystem to $A\cup B$. Again, a Wilson loop threads through all interfaces.
The boundary state may then be expressed as \be
\ket\B = \sum_a\psi_a\bigotimes_{i=1}^{M}\vert\h_a^{i}\dra\,,
\ee
from which one obtains the reduced density matrix $\rho_{AB}$
\begin{align}
&\rho_{AB}=\notag\\
&\quad\sum_a\vert\psi_a\vert^2 \bigotimes_{\Gamma_i=\{\Gamma_{AB}\}}\rho_{AB,a}^{\Gamma_i}\bigotimes_{\Gamma_j=\{\Gamma_{A}\}}\rho_{A,a}^{\Gamma_j}\bigotimes_{\Gamma_k=\{\Gamma_{B}\}}\rho_{B,a}^{\Gamma_k}\,,\quad\lb{rhoMulti}
\end{align}
where we defined
\begin{align}
&\rho_{AB,a}^{\Gamma_i}= \frac{1}{\n_a^i}\sum_{N_i}\sum_{N_i'}e^{-\frac{4\pi \eps}{\ell_i}(h_a+N_i-\frac{c}{24})}e^{-\frac{4\pi \eps}{\ell_i}(h_a+N_i'-\frac{c}{24})}\notag\\
&\hspace{2.4cm}\times\ket{h_a,N_i}\ket{\overline{h_a,N_i}}\bra{h_a,N_i'}\bra{\overline{h_a,N_i'}}\,,\notag\\
&\rho_{A/B,a}^{\Gamma_j} =\frac{1}{\n_a^j}\sum_{N_j}e^{-\frac{8\pi \eps}{\ell_j}(h_a+N_j-\frac{c}{24})}\ket{h_a,N_j}\bra{h_a,N_j}\,.\nn\\[-.6\baselineskip]
\end{align}
Note that the Ishibashi basis vectors appearing in the expression of $\rho_{A/B,a}^{\Gamma_i}$ may be either right or left, depending on the convention, which has no influence on the result.
From the previous cases, the trace of the $n^{\rm th}$ power of the reduced density matrix $\rho_{AA^*}^{(m)}$ for the purification in $\H_A\otimes \H_B\otimes\H_{A^*}\otimes \H_{B^*}$ is straightforward to compute,
\begin{align}
&\hspace{-7pt}\tr\big(\rho_{AA^*}^{(m)}\big)^n =\sum_a\vert\psi_a\vert^{2nm}\prod_{i=\{\Gamma_{AB}\}}\frac{\big(\chi_{h_a}\big( e^{-\frac{8\pi n \eps}{\ell_i}}\big)\big)^2}{\big(\chi_{h_a}\big( e^{-\frac{8\pi \eps}{\ell_i}}\big)\big)^{2n}}\nn\\
&\hspace{1.9cm}\times\prod_{j=\{\Gamma_{A}\cup\Gamma_{B}\}}\frac{\big(\chi_{h_a}\big( e^{-\frac{8\pi m \eps}{\ell_j}}\big)\big)^n}{\big(\chi_{h_a}\big( e^{-\frac{8\pi \eps}{\ell_j}}\big)\big)^{nm}}\,,\nonumber\\
&\hspace{-0.1cm}\simeq e^{\frac{\pi c \ell_{AB}}{24\eps}(\frac{1}{n}-n)}e^{\frac{\pi c(\ell_A+\ell_B)}{48\eps}(\frac{n}{m}-nm)} \nn\\
&\quad\times\sum_a\vert\psi_a\vert^{2nm} (\S_{a0})^{M_{AB}(2(1-n)-n(1-m))+Mn(1-m)}\,,\nn\\[-.6\baselineskip]
\end{align}
where $\ell_{AB}$ and $\ell_{A(B)}$ represent the total length of the interfaces shared between $A$ and $B$, and between $A(B)$ and $C$, respectively. Finally, we obtain the R\'enyi and von Neumann reflected entropies as
\begin{align}
&S_R^{(n)}(A:B) \notag\\
&\hspace{-0.1cm} =\Big(1+\frac{1}{n}\Big) \frac{\pi c}{24}\frac{\ell_{AB}}\eps + \frac{1}{1-n}\log\sum_a \vert\psi_a\vert^{2n} (\S_{a0})^{2M_{AB}(1-n)},\notag\\[-.6\baselineskip]\\
&S_R(A:B) \notag\\ 
&\hspace{-0.1cm}= \frac{\pi c}{12}\frac{\ell_{AB}}\eps +2M_{AB}\sum_a \vert\psi_a\vert^{2} \log\S_{a0} -\sum_a \vert\psi_a\vert^{2} \log\vert\psi_a\vert^{2}\,.\nn\\[-.6\baselineskip] \lb{SRmulti}
\end{align}
We thus find that the (Renyi) reflected entropy depends in general on both the choice of ground state and the elements $\S_{a0}$ of the modular $\S$ matrix. For $M_{AB}=0$ and $M_{AB}=1$, we recover the results of the previous sections, that is for $A$ and $B$ disjoints ($\ell_{AB}=0$) and when $A$ shares only one interface with $B$, respectively. We note that for $A$ and $B$ disjoint with an arbitrary number of components, the (R\'enyi) reflected entropy only depends on the choice of ground state through the Shannon entropy term.

Let us now compare our results with the (R\'enyi) mutual information between $A$ and $B$, which is found to be (see \eqref{MITG} in Appendix \ref{Apdx1})
\begin{align}
&I^{(n)}(A:B) =\Big(1+\frac{1}{n}\Big) \frac{\pi c}{24}\frac{\ell_{AB}}\eps \nn\\
&\hspace{1.2cm}+ \frac{1}{1-n}\log\frac{\sum_a \vert\psi_a\vert^{2n} (\S_{a0})^{(M_{AB}+M_A)(1-n)}}{\sum_a \vert\psi_a\vert^{2n} (\S_{a0})^{(M_{A}+M_B)(1-n)}}\notag\\
&\hspace{1.2cm}+ \frac{1}{1-n}\log\sum_a \vert\psi_a\vert^{2n} (\S_{a0})^{(M_{AB}+M_B)(1-n)},\notag\\[-.6\baselineskip]\\
&I(A:B) \notag\\ 
&\hspace{0.cm}= \frac{\pi c}{12}\frac{\ell_{AB}}\eps +2M_{AB}\sum_a \vert\psi_a\vert^{2} \log\S_{a0} -\sum_a \vert\psi_a\vert^{2} \log\vert\psi_a\vert^{2}\,,\nn\\
&=S_R(A:B)\,.
\end{align}
We observe that the reflected entropy agrees with the mutual information. This is not the case in general for their R\'enyi $n>1$ versions. Only for $M_A=M_B=M_{AB}$ the R\'enyi reflected entropy equals the R\'enyi mutual information.

\subsubsection{Two disjoint regions with contractible $A$ and non-contractible $B$}
So far, we have only considered non-contractible regions on the 2-torus. 
Let us now compute the reflected entropy of two disjoint regions with $A$ contractible and $B$ non-contractible, with a non-contractible complementary region $C$. The geometry can be seen in Fig.\,\hyperref[fig2]{\ref{fig2}(c)}. A Wilson loop threads only through the interfaces between $B$ and $C$, i.e.\;through $\Gamma_2$ and $\Gamma_3$.
The boundary state may thus be expressed as
\be
\ket\B = \vert\h_I^1\dra \otimes\sum_a\psi_a\vert\h_a^{2}\dra\otimes\vert\h_a^3\dra\,,
\ee
where $I$ is the identity topological sector. The reduced density matrix $\rho_{AB}$ then reads
\begin{align}
\rho_{AB}=\rho_{A,I}^{\Gamma_1}\otimes\sum_{a}\vert\psi_a\vert^2\rho_{B,a}^{\Gamma_2}\otimes\rho_{B,a}^{\Gamma_3}\,,\lb{rhoAcontr}
\end{align}
where
\begin{align}
\rho_{A,I}^{\Gamma_1} &=\frac{1}{\n_I^1}\sum_{N_1}e^{-\frac{8\pi \eps}{\ell_1}(h_I+N_1-\frac{c}{24})}\ket{h_I,N_1}\bra{h_I,N_1}\,,\nn\\
\rho_{B,a}^{\Gamma_i} &=\frac{1}{\n_a^i}\sum_{N_i}e^{-\frac{8\pi \eps}{\ell_i}(h_a+N_i-\frac{c}{24})}\ket{\overline{h_a,N_i}}\bra{\overline{h_a,N_i}}\,.\nn\\[-.7\baselineskip]
\end{align}
The purification $\ket{\rho_{AB}^{m/2}}\in\H_A\otimes \H_B\otimes\H_{A^*}\otimes \H_{B^*}$ and its associated reduced density matrix for the subsystem $AA^*$ are obtained through the same procedure as before. One gets
\begin{align}
\tr\big(\rho_{AA^*}^{(m)}\big)^n &\nn\\
&\hspace{-1.72cm}=\frac{\big(\chi_{h_I}\big( e^{-\frac{8\pi m \eps}{\ell_1}}\big)\big)^n}{\big(\chi_{h_I}\big( e^{-\frac{8\pi \eps}{\ell_1}}\big)\big)^{nm}}\left[\sum_a\vert\psi_a\vert^{2m}\prod_{i=2,3}\frac{\chi_{h_a}\big( e^{-\frac{8\pi m \eps}{\ell_i}}\big)}{\big(\chi_{h_a}\big( e^{-\frac{8\pi \eps}{\ell_i}}\big)\big)^{m}}\right]^n\nonumber\\
&\hspace{-1.72cm}\simeq e^{\frac{\pi c(\ell_1+\ell_2+\ell_3)}{48\eps}(\frac{n}{m}-nm)} \S_{00}^{n(1-m)}\Big(\sum_a\vert\psi_a\vert^{2m} (\S_{a0})^{2(1-m)}\Big)^n.\nn\\[-.3\baselineskip]\lb{trAcontr}
\end{align}
The (R\'enyi) reflected entropy thus identically vanishes, $S_R^{(n)}(A:B)=0$. This should have been expected since $\rho_{AB}$ is a factorized state. One can easily show that the R\'enyi mutual information for the configuration in Fig.\,\hyperref[fig2]{\ref{fig2}(c)} also vanishes (see \eqref{MITcA} in Appendix \ref{Apdx1}), hence $S_R^{(n)}(A:B)=0=I^{(n)}(A:B)$.

\subsubsection{Two adjacent non-contractible regions with \mbox{contractible $C$}}
Our last case of interest is that of two adjacent non-contractible regions $A$ and $B$ with a contractible region $C$, as shown in Fig.\,\hyperref[fig2]{\ref{fig2}(d)}. A Wilson loop threads only through the interfaces between $A$ and $B$, i.e. $\Gamma_3$ and $\Gamma_4$.
Similar to the previous case, the boundary state may be expressed as  
\be
\ket\B = \vert\h_I^1\dra \otimes\vert\h_I^2\dra \otimes\sum_a\psi_a\vert\h_a^{3}\dra\otimes\vert\h_a^4\dra\,,
\ee
where $I$ is the identity topological sector. It is straightforward to check that
\begin{align}
\rho_{AB}=\rho_{A,I}^{\Gamma_1}\otimes\rho_{B,I}^{\Gamma_2}\otimes\sum_{a,a'}\psi_a\psi_{a'}^* \vert\h_a^{3}\dra\dla\h_{a'}^{3}\vert\otimes\vert\h_a^{4}\dra\dla\h_{a'}^{4}\vert\,,\quad\nn\\[-.6\baselineskip]\lb{rhoCcontr}
\end{align}
where
\begin{align}
\rho_{A,I}^{\Gamma_1} &=\frac{1}{\n_I^1}\sum_{N_1}e^{-\frac{8\pi \eps}{\ell_1}(h_I+N_1-\frac{c}{24})}\ket{h_I,N_1}\bra{h_I,N_1}\,,\nn\\
\rho_{B,I}^{\Gamma_2} &=\frac{1}{\n_I^2}\sum_{N_2}e^{-\frac{8\pi \eps}{\ell_2}(h_I+N_2-\frac{c}{24})}\ket{\overline{h_I,N_2}}\bra{\overline{h_I,N_2}}\,.\nn\\[-.8\baselineskip]
\end{align}

\vspace{-.13cm}

\noindent The by-now familiar procedure to construct the purification $\ket{\rho_{AB}^{m/2}}$ and the reduced density matrix $\rho_{AA^*}^{(m)}$ yields
\begin{align}
&\tr\big(\rho_{AA^*}^{(m)}\big)^n \nn\\
&=\prod_{i=1}^2\frac{\big(\chi_{h_I}\big( e^{-\frac{8\pi m \eps}{\ell_i}}\big)\big)^n}{\big(\chi_{h_I}\big( e^{-\frac{8\pi \eps}{\ell_i}}\big)\big)^{nm}}\hspace{-3pt}\left(\hspace{-2pt}\sum_a\vert\psi_a\vert^{2n}\prod_{j=3}^4\frac{\chi_{h_a}\big( e^{-\frac{8\pi n \eps}{\ell_j}}\big)}{\big(\chi_{h_a}\big( e^{-\frac{8\pi \eps}{\ell_j}}\big)\big)^{n}}\right)^{\hspace{-3pt}2}\hspace{-3pt},\nonumber\\
&\simeq e^{\frac{\pi c(\ell_1+\ell_2)}{48\eps}(\frac{n}{m}-nm)} e^{\frac{\pi c (\ell_3+\ell_4)}{24\eps}(\frac{1}{n}-n)}\nn\\
&\hspace{2cm}\times\S_{00}^{2n(1-m)}\Big(\sum_a\vert\psi_a\vert^{2n} (\S_{a0})^{2(1-n)}\Big)^2.\nn\\[-.5\baselineskip]
\end{align}
The R\'enyi and von Neumann reflected entropies then read
\begin{align}
&S_R^{(n)}(A:B) \notag\\
&\hspace{-0.cm} =\Big(1+\frac{1}{n}\Big) \frac{\pi c}{24}\frac{\ell_3+\ell_4}\eps + \frac{2}{1-n}\log\sum_a \vert\psi_a\vert^{2n} (\S_{a0})^{2(1-n)},\notag\\[-.6\baselineskip]\lb{SRCcontr}\\
&S_R(A:B) \notag\\ 
&\hspace{0.cm}= \frac{\pi c}{12}\frac{\ell_3+\ell_4}\eps +4\sum_a \vert\psi_a\vert^{2} \log\S_{a0}-2\sum_a \vert\psi_a\vert^{2} \log\vert\psi_a\vert^{2}.\nn\\[-.6\baselineskip]
\end{align}
The mutual information corresponding to the configuration in Fig.\,\hyperref[fig2]{\ref{fig2}(d)} can be found in \cite{Wen:2016snr} (see \eqref{MITcC} in \mbox{Appendix} \ref{Apdx1}). We have that \mbox{$S_R^{(n)}(A:B)=I^{(n)}(A:B)$}. Notice that setting $C$ empty does not change the reflected entropy/mutual information which actually corresponds to twice the bipartite entanglement entropy.

\section{Reflected entropy via surgery} \lb{sec3}

In this section, we compute the topological reflected entropy from a bulk perspective using the surgery method \cite{Witten:1988hf,Witten:1991mm,Dong:2008ft,Wen:2016bla}. We adopt the approach of \cite{Dong:2008ft} which uses a formal description of TQFT, and therefore only computes (universal) finite corrections to the area-law terms.
The evaluation of partition functions on \mbox{various} three-manifolds can be achieved systematically by surgery operations \cite{Witten:1988hf}. These partition functions are related to certain elements of the modular matrix $\S$. For example, the Chern-Simons partition function on $S^3$ with a Wilson loop in representation $R_a$ is given by
\be
Z(S^3, R_a) = \S_{a0}\,,
\ee
while on $S^2\times S^1$, i.e. two solid tori $D_2\times S^1$ glued along their boundaries with Wilson loops in representation $R_a$ and $R_b$ respectively, the partition function reads
\be
Z(S^2\times S^1, R_a,R_b) = \delta_{ab}\,.
\ee 
We will also rely on the basic result that applies to a three-manifold $\M$ which is the connected
sum of two three-manifolds $\M_1$ and $\M_2$ joined along an $S^2$ \cite{Witten:1988hf}:
\be
Z(\M)\times Z(S^3) = Z(\M_1)\times Z(\M_2)\,.\lb{witten}
\ee
The relation \eqref{witten} extends straightforwardly to $\M_1$ and $\M_2$ joined along $n$ $S^2$'s, i.e.
\be
Z(\M) = \frac{Z(\M_1)\times Z(\M_2)}{Z(S^3)^n}\,.
\ee
It is noted that in our discussions, a spatial manifold is two-dimensional and can be viewed as the
boundary of the three-dimensional spacetime manifold where the state is defined.

%
%
%
\begin{widetext}
\begin{figure*}[t]
\centering\hspace*{-6pt}
\includegraphics[scale=0.825]{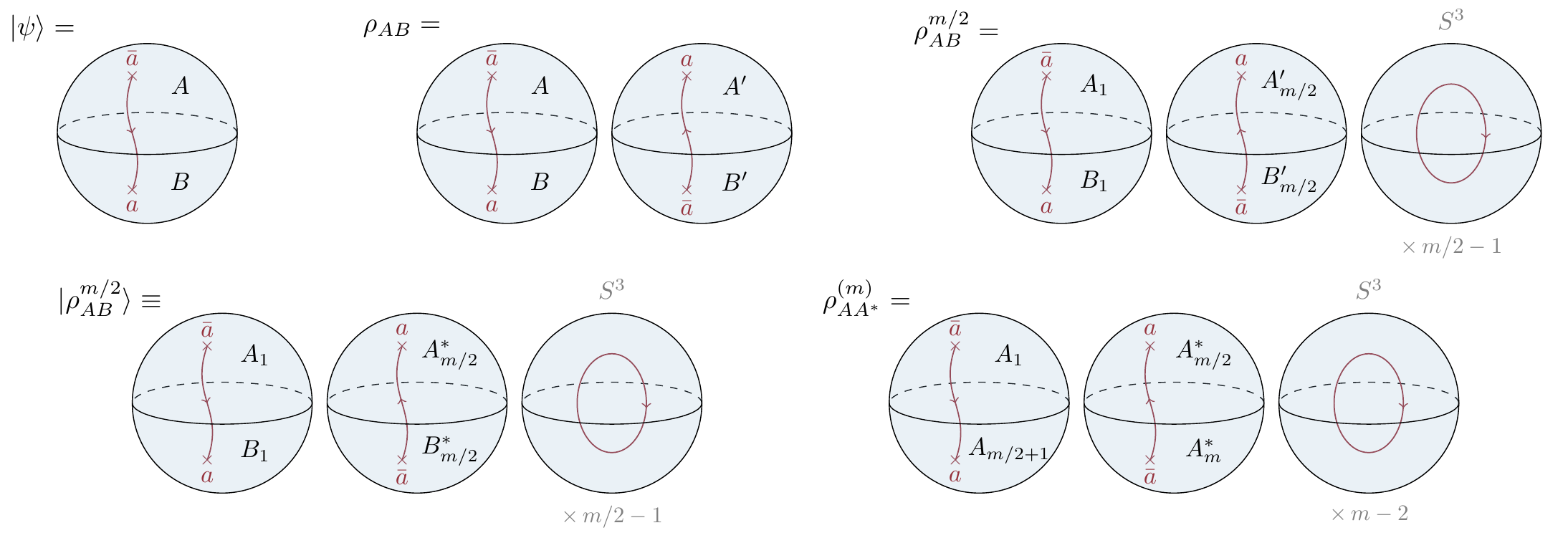}
\caption{Surgery: Pure bipartite state on the 2-sphere. The density matrix $\rho_{AB}$ is two 3-balls. The gluing of $m/2$ density matrices results in two 3-balls and $m/2-1$ $S^3$'s. The canonical purification $\vert \rho_{AB}^{m/2}\ra$ is obtained by interpreting the second 3-ball in $\rho_{AB}^{m/2}$ as living on $\H_{A^*}\otimes\H_{B^*}$. Finally, from $\vert\rho_{AB}^{(m)}\ra$ one gets the reduced density matrix $\rho_{AA^*}^{(m)}$ by tracing out $BB^*$.}
   \label{fig3}
\end{figure*}
\end{widetext}
%
%
To compute the reflected entropy between two subsystems $A$ and $B$, we use the replica trick and proceed as follows. First, we choose our bulk ground state $\ket\psi$ and compute the (reduced) density matrix $\rho_{AB}$. 
Next, we
 glue $m/2$ copies of $\rho_{AB}$ together and construct the purification $\ket{\rho_{AB}^{m/2}}$ by a canonical duplication of the Hilbert space. We then compute $\rho_{AA^*}^{(m)}=\tr_{BB^*}(\ket{\rho_{AB}^{m/2}}\bra{\rho_{AB}^{m/2}})$ and glue $n$ copies of $\rho_{AA^*}^{(m)}$ as to obtain $\tr\big(\rho_{AA^*}^{(m)}\big)^n$.

\subsection{Sphere}
We begin with the simplest case in which the spatial manifold is a 2-sphere $S^2$. In all configurations we study, there are two conjugate quasi-particles on the sphere, one in each regions $A$ and $B$, connected by a Wilson line in a definite representation $R_a$ that thread through all interfaces. One should thus set $\vert\psi_{a'}\vert^2=\delta_{aa'}$ in the formulae obtained using the edge theory approach.

\subsubsection{Pure state}

For pedagogical purpose, let us first consider the topological reflected entropy of a bipartite pure state, which we know should be twice the topological entanglement entropy.
The wave function $\ket{\psi}$ under consideration is a 3-ball depicted in Fig.\,\ref{fig3}. The density matrix \mbox{$\rho_{AB}=\vert\psi\ra\la\psi\vert$} is two 3-balls with conjugate punctures (there is no partial trace to be taken here). To define $\vert\rho_{AB}^{m/2}\ra$ who lives in the doubled Hilbert space $(\H_A\otimes\H_B)\otimes(\H_{A^*}\otimes\H_{B^*})$, we need to compute $\rho_{AB}^{m/2}$, where $m\in2\mathbb{Z}^+$. This is achieved by gluing the region $A'$ ($B'$) in the $i$-th copy of $\rho_{AB}$ to the region $A$ ($B$) in the $(i+1)$-th copy, for $i=1,\cdots,m/2-1$. We are left with two 3-balls from the first and $m/2$-th copies of the density matrix, and $m/2-1$ $S^3$ from sewing the copies together. Then $\rho_{AB}^{m/2}$ on $\H_A\otimes\H_B$ is interpreted as the pure state $\vert\rho_{AB}^{m/2}\ra$ on $\H_A\otimes\H_B\otimes\H_{A^*}\otimes\H_{B^*}$ as suggested in Fig.\,\ref{fig3}, i.e.\;we consider that the second 3-ball in $\rho_{AB}^{m/2}$ lives on $\H_{A^*}\otimes\H_{B^*}$. Next, we construct the reduced density matrix $\rho_{AA^*}^{(m)}$ by tracing over the regions $B$ and $B^*$, which is shown in Fig.\,\ref{fig3}. 
Finally, we can compute $\tr\big(\rho_{AA^*}^{(m)}\big)^n$. We take $n$ copies of $\rho_{AA^*}^{(m)}$ and glue the region $A$ ($A^*$) in the $j$-th copy to the region $A$ ($B$) in the \mbox{$(j+1)$-th} (mod $n$) copy. The resulting manifold is composed of $nm+2(1-n)$ independent $S^3$'s. 
We thus have 
\begin{align}
\frac{\tr\big(\rho_{AA^*}^{(m)}\big)^n}{\big(\Tr\rho_{AB}^m\big)^n} &= \frac{Z(S^3, R_a)^{2(1-n)+nm}}{Z(S^3, R_a)^{nm}}\nn\\
& = Z(S^3, R_a)^{2(1-n)} = \S_{a0}^{2(1-n)}\,,
\end{align}
from which we obtain the (R\'enyi) topological reflected entropy
\be
S_{R,topo}^{(n)}(A:B) = 2\log \S_{a0}= 2S_{topo}^{(n)}(A)\,,
\ee
in agreement with \eqref{LREE} and \eqref{pureS}.

\subsubsection{Two disjoint regions}
We now consider the more interesting case of a tripartite spatial manifold $S^2$ with $A$ and $B$ separated by $C$, as shown\footnote{All the figures of the surgery operations subsequently referred to are gathered in Appendix \ref{Apdx2}.} in Fig.\,\ref{fig4}. First, the wave function manifold is deformed into a topologically equivalent one, that is two 3-balls connected by a tube, as depicted in Fig.\,\ref{fig4}. 
The reduced density matrix $\rho_{AB}$ can then be obtained by tracing over $C$, see again Fig.\,\ref{fig4}. Next, we glue $m/2$ reduced density matrices $\rho_{AB}$ and define the purification $\ket{\rho_{AB}^{m/2}}$ analogous to the previous case, that is by interpreting the region $A'\,(B')$  in the $m/2$-th copy of $\rho_{AB}$ as $A^*\,(B^*)$. After that we get $\rho_{AA^*}^{(m)}$ by tracing over $B$ and $B^*$. Finally, taking $n$ copies of $\rho_{AA^*}^{(m)}$ and gluing the regions $A$ and $A^*$ of each copies cyclically, we obtain $\tr\big(\rho_{AA^*}^{(m)}\big)^n$. The resulting manifold is $2n$ $S^3$'s connected by $nm$ tubes, as illustrated for $m=4$ and $n=2$ in Fig.\,\ref{fig4}. By cutting each tubes and using \eqref{witten}, we find
\begin{align}
\frac{\tr\big(\rho_{AA^*}^{(m)}\big)^n}{\big(\Tr\rho_{AB}^m\big)^n} &= \frac{Z(S^3, R_a)^{n(2-m)}}{Z(S^3, R_a)^{n(2-m)}}=1\,.
\end{align}
The topological reflected entropy thus vanishes for two disjoint regions,
as expected from \eqref{disjointSRS} since $\psi_{a'}=\delta_{aa'}$ here.

\subsubsection{Two adjacent regions}

Finally, we discuss the situation of two adjacent regions $A$ and $B$ on the 2-sphere, depicted in Fig.\,\hyperref[fig5]{\ref{fig5}}. As we did for the disjoint configuration, we start by deforming the three-manifold into two 3-balls connected by a tube. The reduced density matrix is equivalent to three 3-balls connected by two tubes, and the procedure yielding the purification $\ket{\rho_{AB}^{m/2}}$ and the corresponding reduced density matrix $\rho_{AA^*}^{(m)}$ is the same as for the previous cases. Then, gluing together $n$ copies of $\rho_{AA^*}^{(m)}$, we obtain $\tr\big(\rho_{AA^*}^{(m)}\big)^n$. As illustrated in Fig.\,\hyperref[fig5]{\ref{fig5}} for $m=4$ and $n=2$, the resulting manifold is $2+n(m-1)$ $S^3$'s connected by $nm$ tubes. After surgically removing the tubes, we obtain
\begin{align}
   \frac{\tr\big(\rho_{AA^*}^{(m)}\big)^n}{\big(\Tr\rho_{AB}^m\big)^n} & = \frac{Z(S^3, R_a)^{2-n}}{Z(S^3, R_a)^{n}}\nn\\
   &=Z(S^3, R_a)^{2(1-n)}= \S_{a0}^{2(1-n)}\,,
\end{align}
and we recover \eqref{sphere-adjacent}, remembering that the surgery was carried out for a ground state with a Wilson line in a definite topological sector $a$ such that $\psi_{a'}=\delta_{aa'}$ in \eqref{sphere-adjacent}.

\subsection{Torus}

We now focus on a manifold with non-vanishing genus, namely the 2-torus with a Wilson loop in representation $R_a$ present along its center. A solid torus can be thought of as $D_2\times S^1$, and two copies glued together is an $S^2\times S^1$.

\subsubsection{Pure state}

As a warm-up, we first consider a bipartite slicing of the torus into non-contractible $A$ and $B$ regions, as illustrated in Fig.\,\hyperref[fig6]{\ref{fig6}}. For this pure state configuration, the topological reflected entropy should be twice the topological entanglement entropy. 
In a similar manner as done for the sphere, to compute the reflected entropy on the torus, one may deform the manifold into topological equivalent ones which are easier to handle during the surgery procedures.
It is convenient to think of the solid torus as two 3-balls connected by two tubes, as shown in Fig.\,\hyperref[fig6]{\ref{fig6}}. To obtain the purified state in the doubled Hilbert space and compute the moments of the associated reduced density matrix $\tr\big(\rho_{AA^*}^{(m)}\big)^n$, we follow the procedure described in the previous section, see also Fig.\,\hyperref[fig6]{\ref{fig6}}. 
We find that $\tr\big(\rho_{AA^*}^{(m)}\big)^n$ is composed of $n(m-2)$ $S^2\times S^1$'s and two pairs of $S^3$'s joined along $2n$ tubes. Note that $\Tr\rho_{AB}^m$ is $m$ independent $S^2\times S^1$'s.
We thus have
\begin{align}
   \frac{\tr\big(\rho_{AA^*}^{(m)}\big)^n}{\big(\Tr\rho_{AB}^m\big)^n} & =
   \frac{Z(S^3, R_a)^{4(1-n)}Z(S^2\times S^1, R_a,   \bar R_a)^{n(m-2)}}{Z(S^2\times S^1,R_a, \bar R_a)^{nm}}\nn\\
   &=\S_{a0}^{4(1-n)}\,,
\end{align}
where we have used the fact that \mbox{$Z(S^2\times S^1,R_a, \bar R_a)=1$.} 
It is then straightforward to show that for a general ground state where the
Wilson loop is in a superposition of different representations $R_a$ (i.e.\,\,no longer in a definite topological sector $a$), the above generalizes to 
\begin{align}
  \frac{\tr\big(\rho_{AA^*}^{(m)}\big)^n}{\big(\Tr\rho_{AB}^m\big)^n} & =
 \frac{\big(\sum_a\vert\psi_a\vert^{2n} (\S_{a0})^{2(1-n)}\big)^2}{\big(\sum_a\vert\psi_a\vert^{2} \big)^{2n}}\,,
\end{align}
such that the topological (R\'enyi) reflected
entropy is twice the topological (R\'enyi) entanglement entropy computed in \cite{Dong:2008ft},
$S_{R,topo}^{(n)}(A:B) = 2S_{topo}^{(n)}(A)$,
as expected for a pure state.

\subsubsection{Two disjoint non-contractible regions}
For two non-contractible disjoint regions on the torus, the manifold is equivalent to four 3-balls connected by four tubes as depicted in Fig.\,\ref{fig7}. 
Tracing over $C$, we obtain the reduced density matrix $\rho_{AB}$ as four 3-balls joined by five tubes. Then we compute $\rho_{AB}^{m/2}$ and canonically duplicate the Hilbert space to obtain the purified state $\vert\rho_{AB}^{m/2}\ra$ and the associated reduced density matrix $\rho_{AA^*}^{(m)}$. The calculation of $\tr\big(\rho_{AA^*}^{(m)}\big)^n$ results in a manifold of $4n$ $S^3$'s connected by $4nm$ tubes, which after surgically removing the tubes yields
\begin{align}
\frac{\tr\big(\rho_{AA^*}^{(m)}\big)^n}{\big(\Tr\rho_{AB}^m\big)^n} &= \frac{Z(S^3, R_a)^{4n(1-m)}}{Z(S^3, R_a)^{4n(1-m)}}=1\,.
\end{align}
The manifold corresponding to $\tr\big(\rho_{AA^*}^{(m)}\big)^n$ for $m=4$ and $n=2$ is depicted in Fig.\,\ref{fig7}. For a general state in a non-definite topological sector, the above generalizes to
\begin{align}
\frac{\tr\big(\rho_{AA^*}^{(m)}\big)^n}{\big(\Tr\rho_{AB}^m\big)^n} &= \frac{\sum_a\vert\psi_a\vert^{2nm} (\S_{a0})^{4n(1-m)}}{\big(\sum_a\vert\psi_a\vert^{2m} (\S_{a0})^{4(1-m)}\big)^n}\,, 
\end{align}
and we recover the reflected entropy \eqref{SRTdis}.

\subsubsection{Two adjacent non-contractible regions}

Next, we consider two non-contractible adjacent regions on the torus, which is topologically equivalent to three 3-balls connected by three tubes, see
Fig.\,\hyperref[fig9]{\ref{fig9}}. 
The reduced density matrix $\rho_{AB}$ is four 3-balls joined by five tubes. We may then proceed as previously to compute $\tr\big(\rho_{AA^*}^{(m)}\big)^n$, finding that the resulting manifold is $2+nm$ $S^3$'s joined by $3nm$ tubes. We thus obtain
\begin{align}
\frac{\tr\big(\rho_{AA^*}^{(m)}\big)^n}{\big(\Tr\rho_{AB}^m\big)^n} &= \frac{Z(S^3, R_a)^{2(1-nm)}}{Z(S^3, R_a)^{2n(1-m)}}=S_{a0}^{2(1-n)}\,.
\end{align}
 The manifold corresponding to $\tr\big(\rho_{AA^*}^{(m)}\big)^n$ is quite intricate, as illustrated in Fig.\,\hyperref[fig9]{\ref{fig9}} for $m=4$ and $n=2$.
For a general ground state one has
\begin{align}
   \frac{\tr\big(\rho_{AA^*}^{(m)}\big)^n}{\big(\Tr\rho_{AB}^m\big)^n} & =
   \frac{\sum_a\vert\psi_a\vert^{2nm} (\S_{a0})^{2(1-nm)}}{\big(\sum_a\vert\psi_a\vert^{2m} (\S_{a0})^{2(1-m)}\big)^n}\,,
\end{align}
which gives the R\'enyi reflected entropy \eqref{SRRadj}.

\subsubsection{Two disjoint regions with contractible $A$ and non-contractible $B$}

When one of the two disjoint regions is contractible, say $A$ as shown in Fig.\,\hyperref[fig8]{\ref{fig8}}, the solid torus is deformed into three 3-balls connected by three tubes with a Wilson loop that threads only through regions $B$ and $C$. Following the same procedure as before, the calculation of $\tr\big(\rho_{AA^*}^{(m)}\big)^n$ by surgery yields $3n$ $S^3$'s joined along $3nm$ tubes. Note however that the Wilson lines do not thread through every 3-spheres and tubes. Indeed, as illustrated in Fig.\,\hyperref[fig8]{\ref{fig8}} for $m=4$ and $n=2$, there are $n$ $S^3$'s connected by $nm$ tubes that do not contain any Wilson lines, while $n$ pairs of $S^3$'s joined along $2m$ tubes (for each pair) contain some. A 3-sphere with no Wilson line threading through it contributes a $Z(S^3)\equiv Z(S^3,R_0)$ after the surgery.
Removing the tubes and applying \eqref{witten} we thus obtain
\begin{align}
   \frac{\tr\big(\rho_{AA^*}^{(m)}\big)^n}{\big(\Tr\rho_{AB}^m\big)^n} & = \frac{Z(S^3, R_a)^{2n(1-m)}Z(S^3)^{n(1-m)}}{Z(S^3, R_a)^{2n(1-m)}Z(S^3)^{n(1-m)}}=1\,.\quad
\end{align}
For a general ground state, the result becomes
\begin{align}
   \frac{\tr\big(\rho_{AA^*}^{(m)}\big)^n}{\big(\Tr\rho_{AB}^m\big)^n} & =
   \frac{\big(\sum_a\vert\psi_a\vert^{2m} (\S_{a0})^{2(1-m)}\big)^n(\S_{00})^{n(1-m)}}{\big(\sum_a\vert\psi_a\vert^{2m} (\S_{a0})^{2(1-m)}(\S_{00})^{1-m}\big)^n}\,.
\end{align}
Hence $S_R^{(n)}(A:B) =0$ in agreement the with edge theory calculation.

\subsubsection{Two adjacent non-contractible regions with \mbox{contractible $C$}}

Finally, we consider two non-contractible adjacent regions $A$ and $B$ on the 2-torus with a contractible region $C$. The manifold is first deformed into four 3-balls connected by four tubes, as one can see in Fig.\,\hyperref[fig10]{\ref{fig10}}, where the Wilson loop only pierces through the interfaces between $A$ and $B$. Then following the familiar canonical purification procedure, the surgery yields $\tr\big(\rho_{AA^*}^{(m)}\big)^n$ as a manifold of $2n(m-1)+4$ $S^3$'s joined along $4nm$ tubes. The different 3-spheres and tubes are arranged in a complex way, as one may appreciate for $m=4$ and $n=2$ in Fig.\,\hyperref[fig10]{\ref{fig10}}. 
There are Wilson loops threading through $2n(m-2)$ $S^3$'s connected to $2n(m-2)$ tubes (represented as red tubes in Fig.\,\hyperref[fig10]{\ref{fig10}}) as well as through two pairs of $S^3$'s joined along $2n$ tubes each. The remaining $2n$ $S^3$'s and $2nm$ tubes do not contain any Wilson loops. We thus obtain
\begin{align}
\frac{\tr\big(\rho_{AA^*}^{(m)}\big)^n}{\big(\Tr\rho_{AB}^m\big)^n} & = \frac{Z(S^3, R_a)^{4(1-n)}Z(S^3)^{2n(1-m)}}{Z(S^3)^{2n(1-m)}}\,,\nn\quad\\
&=(\S_{a0})^{4(1-n)}\,.
\end{align}
For a general state, the above generalizes to
\begin{align}
\frac{\tr\big(\rho_{AA^*}^{(m)}\big)^n}{\big(\Tr\rho_{AB}^m\big)^n} & =
\big(\sum_a\vert\psi_a\vert^{2nm} (\S_{a0})^{2(1-n)}\big)^2\,,
\end{align}
which gives the same R\'enyi reflected entropy \eqref{SRCcontr} as with the edge theory approach.

\addtocontents{toc}{\protect\medskip}

\section{Odd entropy} \lb{sec4}
The odd entropy of a bipartite state $\rho_{AB}$, introduced in \cite{Tamaoka:2018ned}, involves an analytic continuation of the odd sequence at $n_o\rightarrow 1$ of the moments of the partial transpose density matrix,
\be
S_o(A:B) &=& \lim_{n_o\rightarrow1} \frac{1}{1-n_o}\log\tr\big(\rho^{T_B}_{AB}\big)^{n_o}\,,
\ee
where $\cdot^{T_B}$ indicates the partial transposition in $\H_B$. The quantity suggested in \cite{Tamaoka:2018ned} as a dual of the entanglement wedge cross-section is actually a `regulated' form of the odd entropy, denoted hereafter $\E_o$, which is the difference between odd entropy and entanglement entropy,
\be
\E_o(A:B)\equiv S_o(A:B) - S(A\cup B)\,.
\ee
As mentioned in the introduction, another correlation measure that possesses a simple holographic dual interpretation as (twice) the entanglement wedge cross-section is the reflected entropy \cite{Dutta:2019gen}. It is thus interesting to see whether reflected entropy and odd entropy are related to each other in Chern-Simons theories,
\be
S_R(A:B) \stackrel{\scalebox{1.05}{?}}{=} 2\E_o(A:B)\,.
\ee
For pure states we can already observe that the relation above holds since the odd entropy reduces to the entanglement entropy, hence 
\be
\E_o(A:B)=S_o(A:B)=S(A)=\frac{1}2 S_R(A:B)\,.\nn
\ee

\subsection{Sphere}

\subsubsection{Two disjoint regions}
For two disjoint regions on the 2-sphere (see Fig.\,\hyperref[fig1]{\ref{fig1}(b)}), the reduced density matrix is given in \eqref{rhoSdisj} and is invariant under partial transposition, $\tr\big(\rho_{AB}^{T_B}\big)^n=\Tr\rho_{AB}^n$, which immediately leads to
\be
S_o(A:B) = S(A\cup B)\,,\lb{odd-inv}
\ee 
where $S(A\cup B)$ can be found in \eqref{RESdisAB} by taking the $n\rightarrow 1$ limit. This yields
\be
\E_o(A:B) = 0\,.
\ee
Note that the invariance under partial transposition of the reduced density matrix implies \eqref{odd-inv} in general.
The reflected entropy has been computed in \eqref{disjointSRS}. Although $\E_o$ is trivial, in anticipation of the adjacent case we may write
\begin{align}
S_R(A:B) = 2\E_o(A:B) - \sum_a \vert\psi_a\vert^{2} \log\vert\psi_a\vert^{2}\,,
\end{align}
with $S_R(A:B) = 2\E_o(A:B)=0$ for a Wilson line in a definite topological sector.

\subsubsection{Two adjacent regions}
The reduced density matrix, see \eqref{rhoSadj}, corresponding to this configuration is not invariant under partial transposition, therefore $\tr\big(\rho^{T_B}_{AB}\big)^{n}$ depends on whether $n$ is even or odd. For $n=n_o$ odd, one finds that 
\begin{align}
&\tr\big(\rho^{T_B}_{AB}\big)^{n_o} \nn\\
&\qquad= \sum_a\vert\psi_a\vert^{2n_o} \frac{\chi_{h_a}\big( e^{-\frac{8\pi n_o \eps}{\ell_1}}\big)}{\big(\chi_{h_a}\big( e^{-\frac{8\pi \eps}{\ell_1}}\big)\big)^{n_o}}\frac{\chi_{h_a}\big( e^{-\frac{8\pi n_o \eps}{\ell_2}}\big)}{\big(\chi_{h_a}\big( e^{-\frac{8\pi \eps}{\ell_2}}\big)\big)^{n_o}} ,\nn\\
&\qquad\simeq e^{\frac{\pi c}{48}\frac{\ell_1+\ell_2}{\eps}(\frac{1}{n_o}-n_o)}\sum_a\vert\psi_a\vert^{2n_o} (\S_{a0})^{2(1-n_o)}\,.\nn\\[-.6\baselineskip]
\end{align}
The odd entropy  then easily follows,
\begin{align}
& S_o(A:B) \nn\\
&\;\;\; = \frac{\pi c}{24}\frac{\ell_1+\ell_2}\eps  +2\sum_a \vert\psi_a\vert^{2} \log\S_{a0} -\sum_a \vert\psi_a\vert^{2} \log\vert\psi_a\vert^{2}.
\end{align}
The reflected entropy can be found in \eqref{SRadj}, and the entanglement entropy $S(A\cup B)$ is of the form \eqref{LREE}; we thus find the following relation 
\begin{align}
S_R(A:B) = 2\E_o(A:B) - \sum_a \vert\psi_a\vert^{2} \log\vert\psi_a\vert^{2}\,,
\end{align}
where
\begin{align}
\E_o(A:B) =  \frac{\pi c}{24}\frac{\ell_1}\eps +\sum_a \vert\psi_a\vert^{2} \log\S_{a0}\,, \lb{EoSa}
\end{align}
Proportional to the length of the interface shared between $A$ and $B$, the first term in \eqref{EoSa} satisfies the area law. The second term constitutes the topological part of the regulated odd entropy. For an Abelian Chern-Simons theory ($d_a=1$ such that $\S_{a0}=1/\mathcal{D}$ for each topological sector $a$), the topological term does not depend on the choice of ground state, while for a non-Abelian Chern-Simons theory ($d_a>1$ such that $\S_{a0}\neq1/\mathcal{D}$ for at least one topological sector), the topological part does depend on the choice of ground state. The regulated odd entropy thus allows us a to distinguish an Abelian theory from a non-Abelian one. Another quantity that can characterize the abelianity of a Chern-Simons theory is the logarithmic negativity \cite{Wen:2016snr,Wen:2016bla} which, interestingly, also involves a partial transposition in its definition. In characterizing topological phases of matter, the partial transposition thus emerges as a crucial tool%
\footnote{We thank Jonah Kudler-Flam for discussions on this point.}
(see also \cite{PhysRevLett.118.216402} for another interesting example).

\subsection{Torus}

\subsubsection{Non-contractible multi-component interfaces}
For a general configuration on the 2-torus, with non-contractible multi-component $A$, $B$ and $C$, the reduced density matrix is given in \eqref{rhoMulti}. The odd moments of the partial transpose of $\rho_{AB}$ are easily found to be
\begin{align}
&\tr\big(\rho^{T_B}_{AB}\big)^{n_o} = \sum_a\vert\psi_a\vert^{2n_o} \prod_{i=1}^M\frac{\chi_{h_a}\big( e^{-\frac{8\pi n_o \eps}{\ell_i}}\big)}{\big(\chi_{h_a}\big( e^{-\frac{8\pi \eps}{\ell_i}}\big)\big)^{n_o}} ,\nn\quad\\
&\quad\simeq e^{\frac{\pi c}{48}\frac{\ell_A+\ell_B+\ell_{AB}}{\eps}(\frac{1}{n_o}-n_o)}\sum_a\vert\psi_a\vert^{2n_o} (\S_{a0})^{M(1-n_o)},\nn\\[-.6\baselineskip]
\end{align}
where we recall that $M=M_A+M_B+M_{AB}$ is the total number of interfaces, and $\ell_{AB}$ and $\ell_{A(B)}$ represent the total length of the interfaces shared between $A$ and $B$, and between $A(B)$ and $C$, respectively. The odd entropy can then be expressed as
\begin{align}
& S_o(A:B) = \frac{\pi c}{24}\frac{\ell_A+\ell_B+\ell_{AB}}\eps +M\sum_a \vert\psi_a\vert^{2} \log\S_{a0} \nn\\
&\hspace{2.5cm} -\sum_a \vert\psi_a\vert^{2} \log\vert\psi_a\vert^{2}.
\end{align}
Using \eqref{SRmulti} and \eqref{REtorusGen}, we again obtain the relation
\begin{align}
S_R(A:B) = 2\E_o(A:B) - \sum_a \vert\psi_a\vert^{2} \log\vert\psi_a\vert^{2}\,,
\end{align}
where
\begin{align}
\E_o(A:B) =  \frac{\pi c}{24}\frac{\ell_{AB}}\eps +M_{AB}\sum_a \vert\psi_a\vert^{2} \log\S_{a0}\,. \lb{EoTa}
\end{align}
Again, we observe that the topological term in \eqref{EoTa} depends on the choice of ground state for non-Abelian theories, but does not for Abelian ones.

\subsubsection{Two disjoint regions with contractible $A$ and non-contractible $B$}
For the geometry shown in Fig.\,\hyperref[fig2]{\ref{fig2}(c)}, the reduced density matrix corresponding to  $A\cup B$ (see \eqref{rhoAcontr}) is invariant under partial transposition, hence the odd entropy coincides with the entanglement entropy for $A\cup B$ given in \eqref{REcontA}. This yields
\be
S_R(A:B) = 2\E_o(A:B) = 0\,.
\ee

\subsubsection{Two adjacent non-contractible regions with \mbox{contractible $C$}}
Given the reduced density matrix $\rho_{AB}$ in \eqref{rhoCcontr}, it is a straightforward matter to compute
\begin{align}
&\tr\big(\rho^{T_B}_{AB}\big)^{n_o} \nn\\
&=\prod_{i=1,2}\hspace{-0pt}\frac{\chi_{h_I}\big( e^{-\frac{8\pi n_o \eps}{\ell_i}}\big)}{\big(\chi_{h_I}\big( e^{-\frac{8\pi \eps}{\ell_i}}\big)\big)^{n_o}}\sum_a\vert\psi_a\vert^{2n_o}\hspace{-3pt}\prod_{j=3,4}\frac{\chi_{h_a}\big( e^{-\frac{8\pi n_o \eps}{\ell_j}}\big)}{\big(\chi_{h_a}\big( e^{-\frac{8\pi \eps}{\ell_j}}\big)\big)^{n_o}},\nonumber\\
&\simeq e^{\frac{\pi c(\ell_1+\ell_2+\ell_3+\ell_4)}{48\eps}(\frac{1}{n_o}-n_o)} \S_{00}^{2(1-n_o)}\sum_a\vert\psi_a\vert^{2n_o} (\S_{a0})^{2(1-n_o)}.\nn\\[-.3\baselineskip]
\end{align}
The odd entropy can then be expressed as
\begin{align}
& S_o(A:B) = \frac{\pi c}{24}\frac{\sum_{i=1}^4\ell_i}\eps + 2\log S_{00}  + 2\sum_a \vert\psi_a\vert^{2} \log\S_{a0}\nn\\
&\hspace{2.5cm}  -\sum_a \vert\psi_a\vert^{2} \log\vert\psi_a\vert^{2}.
\end{align}
With the entanglement entropy for $A\cup B$ obtained by taking the $n\rightarrow 1$ limit in \eqref{RECcontr}, we get
\begin{align}
S_R(A:B) = 2\E_o(A:B) \,.
\end{align}
We note that in this case, the regulated odd entropy depends on the choice of ground state for both Abelian and non-Abelian theories.

 \section{Discussion} \lb{conclu}
 
We studied the reflected entropy in $(2+1)$-dimensional Chern-Simons theories for a class of mixed states obtained by tracing out the degrees of freedom of some subsystem of a tripartite ground state. We mainly focused on spherical and toroidal spatial manifolds. Relying on its replica formulation \cite{Dutta:2019gen}, we employed two different approaches to compute the reflected entropy. The first one, the edge theory approach \cite{Wen:2016snr}, makes use of the bulk-edge correspondence in TQFT, while with the second method the reflected entropy is computed directly using surgery techniques \cite{Witten:1988hf,Witten:1991mm,Dong:2008ft}. Both approaches yield identical results for all cases studied in this work, namely the reflected entropy coincides with the mutual information,
\be
S_R(A:B)=I(A:B)\,, \lb{SReqI}
\ee
regardless of whether the subsystems $A$ and $B$ are adjacent or disjoint. We have noted, though, that their R\'enyi versions do not agree in general. Such a relation can be observed in two-dimensional holographic CFTs when the contribution is universal, as, e.g., for adjacent intervals \cite{Kudler-Flam:2018qjo,Dutta:2019gen}. 
It was also reported in \cite{Kudler-Flam:2020url,Moosa:2020vcs} for global quenches in two-dimensional rational and holographic CFTs.
Equality between reflected entropy and mutual information implies certain structure properties of the tripartite pure states. In a recent work \cite{Zou:2020bly}, the authors considered the quantity $S_R - I$ as a measure of tripartite entanglement (see also \cite{Akers:2019gcv} in the holographic context). 
Tripartite pure states that satisfy $S_R = I$ for $\rho_{AB}$ have been dubbed \textit{sum of triangle states} in  \cite{Zou:2020bly}. $S_R \neq I$ signals irreducible tripartite entanglement. The equality puts constraints on the nature of tripartite entanglement in the states for which it holds, such that, for example, no $W$--like entanglement. However, it does not imply a complete lack of tripartite entanglement in general, as GHZ states do satisfy $S_R=I$. 

From \eqref{SReqI}, we observe that both the lower bound \eqref{bounds} and the polygamy inequality \eqref{polygamy} for the reflected entropy are saturated. Also, the monotonicity of mutual information together with the relation \eqref{SReqI} trivially implies the monotonicity of reflected entropy, \mbox{$S_R(A,B\cup C)\ge S_R(A:B)$}, for the type of mixed states under consideration.
It would be interesting to see if the monotonicity of the reflected entropy holds for generic mixed states, especially since a R\'enyi version ($n>1$) of this inequality was proven in generality in \cite{Dutta:2019gen}.

We also studied the recently introduced odd entropy \cite{Tamaoka:2018ned}, motivated by the fact that its proposed holographic dual interpretation is similar to that of the reflected entropy. The relevant quantity  is a `regulated' odd entropy, given by the difference between the odd entropy and the entanglement entropy, which we denote $\E_o(A:B)$. We found that the reflected entropy and twice the regulated odd entropy match, up to a classical Shannon term,
\be
S_R(A:B)=2\E_o(A:B) + \a H(\{\psi_a\})\,, \lb{SRodd}\quad
\ee
where $H(\{\psi_a\})=-\sum_a \vert\psi_a\vert^{2} \log\vert\psi_a\vert^{2}$ is the Shannon entropy of the classical probability distribution $\{\vert\psi_a\vert^2\}$. The constant $\a$ is zero if $A$ and/or $B$ and/or $C$ is completely contractible such that there is no Wilson line threading the interface between at least two of the three regions, otherwise it is equal to one. Thus, the reflected entropy and the regulated odd entropy possibly differ only by a Shannon term, coming from a Wilson line fluctuating among different topological sectors and tunneling through the interfaces, whose presence indicates in our setup that the three subsystems are all non-contractible. The relation \eqref{SRodd} also suggests that the reflected entropy (or, equivalently, the mutual information) is more sensible to classical correlations than (twice) the regulated odd entropy, as their difference, if non-zero, is classical. Additionally, we found that the regulated odd entropy for two adjacent (non-contractible) regions on the sphere (torus) can be used to distinguish Abelian Chern-Simons theories from non-Abelian ones, in a very similar manner as the logarithmic negativity (see \cite{Wen:2016snr,Wen:2016bla}).

There are several future avenues worth exploring. First, it is not yet clear what the reflected entropy and (regulated) odd entropy exactly measure in general -- see discussions and recent developments on this issue in \cite{Dutta:2019gen,Kusuki:2019rbk,Kusuki:2019evw,Akers:2019gcv,Kudler-Flam:2020url,Mollabashi:2020ifv,Zou:2020bly} -- which needs to be further investigated. An interesting direction would be to study the reflected entropy and the regulated odd entropy for more general mixed states in $3d$ Chern-Simons theories. Though we believe that the mixed states considered in the present work reflect the essential features arising for generic ones, it is an intriguing question whether the reflected entropy and the mutual information in $3d$ Chern-Simons theories coincide in general, and whether the regulated odd entropy is generically related to the reflected entropy as in \eqref{SRodd}.
One could also revisit our analysis of the reflected entropy and the regulated odd entropy in the context of gapped interfaces in both Abelian and non-Abelian Chern-Simons theories, see, e.g., \cite{Fliss:2017wop,Fliss:2020cos}. Finally, it is worth investigating the reflected entropy in other theories, such as, for example, Lifshitz theories. Lifshitz theories are critical non-relativistic quantum field theories exhibiting anisotropic scaling between space and time \cite{Ardonne2004}, and which are known to display similar entanglement properties as topological theories \cite{Fradkin:2006mb,Zhou:2016ykv,Angel-Ramelli:2020wfo}. It would thus be interesting to compute the reflected entropy in such theories to compare to the results in this paper.

\begin{acknowledgments}

It is a pleasure to thank Jonah Kudler-Flam, Shinsei Ryu, and Xueda Wen for interesting discussions and valuable comments on a first version of this manuscript. We also thank Pratik Rath and Yijian Zou for useful discussions.
This work was supported in part by the National \mbox{Natural} Science Foundation of China (NSFC) Grant No. 11335012, No. 11325522 and No. 11735001. C.B is also supported by a Boya Postdoctoral Fellowship at Peking University.

\end{acknowledgments}

\appendix

\section{Mutual information}\lb{Apdx1}

For the sake of being self-contained, we compute here the (R\'enyi) mutual information for the different cases considered in this paper, most of which can be found in \cite{Wen:2016snr}. The R\'enyi mutual information $I^{(n)}(A:B)$ between two subsystems $A$ and $B$  is defined as
\begin{align}
I^{(n)}(A:B) = S^{(n)}(A) + S^{(n)}(B) - S^{(n)}(A\cup B)\,, \quad
\end{align}
where $S^{(n)}(A)$ is the R\'enyi entropy for the subsystem $A$,
\be
S^{(n)}(A)=\frac{1}{1-n}\log\frac{\Tr\rho_A^n}{(\Tr\rho_A)^n}\,,
\ee
and similarly for $B$ and $A\cup B$.
The mutual information is obtained in terms of entanglement entropies by taking the $n\rightarrow 1$ limit
\be
I(A:B) = \lim_{n\rightarrow 1} I^{(n)}(A:B) \,,
\ee
and is a measure of total correlations between $A$ and $B$.

\medskip

\subsection*{1.\quad Sphere}

\subsubsection*{\textit{a.}\quad Two disjoint regions}
This case corresponds to the configuration in Fig.\,\hyperref[fig1]{\ref{fig1}(b)}. Clearly, $S^{(n)}(A)$ and $S^{(n)}(B)$ have the same form, which is given in \eqref{LRRE} by 
\begin{align}
S^{(n)}(A(B))&=\Big(1+\frac{1}{n}\Big) \frac{\pi c}{48}\frac{\ell_{1(2)}}\eps \nn\\
&\hspace{0.5cm} +\frac{1}{1-n}\log\sum_a \vert\psi_a\vert^{2n} (\S_{a0})^{1-n}\,.
\end{align}
Thus we only have to compute $S^{(n)}(A\cup B)$. Actually, we do not need to do much since we have already calculated $\Tr\rho_{AB}^{m}\equiv\Tr\rho_{AA*}^{(m)}$ in \eqref{rhoSdis}, hence
\be
S^{(n)}(A\cup B) &=& \Big(1+\frac{1}{n}\Big) \frac{\pi c}{48}\frac{\ell_1+\ell_2}\eps \nn\\
&&\hspace{0cm}+ \frac{1}{1-n}\log\sum_a \vert\psi_a\vert^{2n}(\S_{a0})^{2(1-n)}\,.\qquad \lb{RESdisAB}
\ee
We obtain the (R\'enyi) mutual information as
\be
I^{(n)}(A:B) &=& \frac{1}{1-n}\log\frac{\big(\sum_a \vert\psi_a\vert^{2n} (\S_{a0})^{1-n}\big)^2}{\sum_a \vert\psi_a\vert^{2n} (\S_{a0})^{2(1-n)}}\,,\;\;\quad\\[.6\baselineskip] 
I(A:B) &=& -\sum_a \vert\psi_a\vert^{2} \log\vert\psi_a\vert^{2}\,. \lb{MISD}
\ee

\medskip

\subsubsection*{\textit{b.}\quad Two adjacent regions}
This case corresponds to the configuration in Fig.\,\hyperref[fig1]{\ref{fig1}(c)}. The R\'enyi mutual information can be inferred from the previous case. Indeed, $S^{(n)}(A)$ and $S^{(n)}(A\cup B)$ have the same form as \eqref{LRRE}, while $S^{(n)}(B)$ is given by \eqref{RESdisAB}. We immediately get
\begin{align}
&I^{(n)}(A:B) \notag\\
&\hspace{0.5cm} =\Big(1+\frac{1}{n}\Big) \frac{\pi c}{24}\frac{\ell_1}\eps + \frac{1}{1-n}\log\sum_a \vert\psi_a\vert^{2n} (\S_{a0})^{2(1-n)}\,,\notag\\[-.6\baselineskip]\\ 
&I(A:B) \notag\\ 
&\hspace{0.5cm}= \frac{\pi c}{12}\frac{\ell_1}\eps +2\sum_a \vert\psi_a\vert^{2} \log\S_{a0} -\sum_a \vert\psi_a\vert^{2} \log\vert\psi_a\vert^{2}\,.\qquad\nonumber\\[-.6\baselineskip]
\end{align}

\subsection*{2.\quad Torus}

\subsubsection*{\textit{a.}\quad Non-contractible multi-component interfaces}

Instead of reproducing the results of \cite{Wen:2016snr} for the mutual information corresponding to the configurations in Figs.\,\hyperref[fig2]{\ref{fig2}(a)} and \hyperref[fig2]{\ref{fig2}(b)}, we compute the (R\'enyi) mutual information for the more general case where $A$ and $B$ are each composed of an arbitrary number of components with an arbitrary number of shared interfaces between them. 
We recall that there are $M$ interfaces $\Gamma_i$ in total, of three types: $M_{AB}$ between $A$ and $B$, $M_A$ between $A$ and $C$, and $M_B$ between $B$ and $C$, where $C$ is the complementary subsystem to $A\cup B$. A Wilson loop threads through all interfaces.

The reduced density matrices for the subsystems $A$, $B$, and $A\cup B$ are given by
\begin{align}
\rho_{A(B)}&=\sum_a\vert\psi_a\vert^2 \bigotimes_{\Gamma_i=\{\Gamma_{AB}\}}\rho_{AB,a}^{\Gamma_i}\bigotimes_{\Gamma_j=\{\Gamma_{A(B)}\}}\rho_{A(B),a}^{\Gamma_j}\,,\quad\nn\\
\rho_{AB}&=\sum_a\vert\psi_a\vert^2 \bigotimes_{\Gamma_i=\{\Gamma_{A}\}}\rho_{A,a}^{\Gamma_i}\bigotimes_{\Gamma_j=\{\Gamma_{B}\}}\rho_{B,a}^{\Gamma_j}\,,\quad
\end{align}
where we defined
\begin{align}
&\rho_{A(B),a}^{\Gamma_i} =\frac{1}{\n_a^i}\sum_{N_i}e^{-\frac{8\pi \eps}{\ell_i}(h_a+N_i-\frac{c}{24})}\ket{h_a,N_i}\bra{h_a,N_i}\,,\nn\\
&\rho_{AB,a}^{\Gamma_j}= \frac{1}{\n_a^j}\sum_{N_j}\sum_{N_j'}e^{-\frac{4\pi \eps}{\ell_j}(h_a+N_j-\frac{c}{24})}e^{-\frac{4\pi \eps}{\ell_j}(h_a+N_j'-\frac{c}{24})}\notag\\
&\hspace{1.8cm}\times\ket{h_a,N_j}\ket{\overline{h_a,N_j}}\bra{h_a,N_j'}\bra{\overline{h_a,N_j'}}\,.
\end{align}
One can then obtain
\begin{align}
\Tr\rho_{A(B)}^n &=\sum_a\vert\psi_a\vert^{2n}\hspace{-4pt}\prod_{i=\{\Gamma_{A(B)}\cup\Gamma_{AB}\}}\frac{\chi_{h_a}\big( e^{-\frac{8\pi n \eps}{\ell_i}}\big)}{\big(\chi_{h_a}\big( e^{-\frac{8\pi \eps}{\ell_i}}\big)\big)^{n}}\,,\nonumber\\
&\simeq e^{\frac{\pi c(\ell_{A(B)}+\ell_{AB})}{48\eps}(\frac{1}{n}-n)}\nn\\
&\hspace{1.5cm}\times\sum_a\vert\psi_a\vert^{2n} (\S_{a0})^{(M_{A(B)}+M_{AB})(1-n)}\,,\nn\\
\Tr\rho_{AB}^n &=\sum_a\vert\psi_a\vert^{2n}\prod_{j=\{\Gamma_{A}\cup\Gamma_{B}\}}\frac{\chi_{h_a}\big( e^{-\frac{8\pi n \eps}{\ell_j}}\big)}{\big(\chi_{h_a}\big( e^{-\frac{8\pi \eps}{\ell_j}}\big)\big)^{n}}\,,\nonumber\\
&\simeq e^{\frac{\pi c(\ell_A+\ell_B)}{48\eps}(\frac{1}{n}-n)} \sum_a\vert\psi_a\vert^{2n} (\S_{a0})^{(M_A+M_B)(1-n)},\qquad\qquad\nn\\[-.6\baselineskip]
\end{align}
where $\ell_{AB}$ and $\ell_{A(B)}$ represent the total length of the interfaces shared between $A$ and $B$, and between $A(B)$ and $C$, respectively. The corresponding R\'enyi entropies read
\begin{align}
S^{(n)}(A(B)) &= \Big(1+\frac{1}{n}\Big) \frac{\pi c}{48}\frac{\ell_{A(B)}+\ell_{AB}}\eps \nn\\
&\hspace{-0.25cm}+ \frac{1}{1-n}\log\sum_a \vert\psi_a\vert^{2n}(\S_{a0})^{(M_{A(B)}+M_{AB})(1-n)},\nn\\
S^{(n)}(A\cup B) &= \Big(1+\frac{1}{n}\Big) \frac{\pi c}{48}\frac{\ell_A+\ell_B}\eps \nn\\
&\hspace{-0.25cm}+ \frac{1}{1-n}\log\sum_a \vert\psi_a\vert^{2n}(\S_{a0})^{(M_{A}+M_{B})(1-n)}\,,\quad\nn\\[-.6\baselineskip]   \lb{REtorusGen}
\end{align}
based on which the (R\'enyi) mutual information between $A$ and $B$ follows
\begin{align}
&I^{(n)}(A:B) =\Big(1+\frac{1}{n}\Big) \frac{\pi c}{24}\frac{\ell_{AB}}\eps \nn\\
&\hspace{1.2cm}+ \frac{1}{1-n}\log\frac{\sum_a \vert\psi_a\vert^{2n} (\S_{a0})^{(M_{AB}+M_A)(1-n)}}{\sum_a \vert\psi_a\vert^{2n} (\S_{a0})^{(M_{A}+M_B)(1-n)}}\notag\\
&\hspace{1.2cm}+ \frac{1}{1-n}\log\sum_a \vert\psi_a\vert^{2n} (\S_{a0})^{(M_{AB}+M_B)(1-n)},\notag\\
&I(A:B) \notag\\ 
&\hspace{-0.08cm}= \frac{\pi c}{12}\frac{\ell_{AB}}\eps +2M_{AB}\sum_a \vert\psi_a\vert^{2} \log\S_{a0} -\sum_a \vert\psi_a\vert^{2} \log\vert\psi_a\vert^{2}\,. \lb{MITG}
\end{align}
The configurations corresponding to Figs.\,\hyperref[fig2]{\ref{fig2}(a)} and \hyperref[fig2]{\ref{fig2}(b)}
are recovered for $M_{AB}=\ell_{AB}=0, M_A=M_B=2$ and $M_{AB}=M_A=M_B=1$, respectively.

\subsubsection*{\textit{b.}\quad Two disjoint regions with contractible $A$ and non-contractible $B$}
For the geometry shown in Fig.\,\hyperref[fig2]{\ref{fig2}(c)}, we can directly get $S^{(n)}(B)$ from \eqref{REtorusGen} by setting $M_B=2$, $\ell_B=\ell_2+\ell_3$ and $\ell_{AB}=0=M_{AB}$. Furthermore, we already obtained $\Tr\rho_{AB}^{m}\equiv\Tr\rho_{AA*}^{(m)}$ in \eqref{trAcontr}, yielding
\be
S^{(n)}(A\cup B) &=& \Big(1+\frac{1}{n}\Big) \frac{\pi c}{48}\frac{\ell_1+\ell_2+\ell_3}\eps  + \log S_{00}\nn\\
&&\hspace{0cm}+ \frac{1}{1-n}\log\sum_a \vert\psi_a\vert^{2n}(\S_{a0})^{2(1-n)}\,.\qquad\quad\lb{REcontA}
\ee
Thus, we only need to compute $S^{(n)}(A)$. The boundary state at the interface $\Gamma_1$ simply is $\ket\B = \vert\h_I^1\dra$, and the corresponding reduced density matrix for $A$ reads
\be
\rho_{A} =\frac{1}{\n_I^1}\sum_{N}e^{-\frac{8\pi \eps}{\ell_1}(h_I+N-\frac{c}{24})}\ket{h_I,N}\bra{h_I,N}\,,\qquad
\ee
which gives the following R\'enyi entropies
\be
S^{(n)}(A) = \Big(1+\frac{1}{n}\Big) \frac{\pi c}{48}\frac{\ell_1}\eps  + \log S_{00}\,.\quad
\ee
It is then straightforward to check that the (R\'enyi) mutual information identically vanishes,
\be
I^{(n)}(A:B) = 0\,. \lb{MITcA}
\ee

\subsubsection*{\textit{c.}\quad Two adjacent non-contractible regions with \mbox{contractible $C$}}
This case, corresponding to the configuration in Fig.\,\hyperref[fig2]{\ref{fig2}(d)}, has been treated in detail in \cite{Wen:2016snr}. Let us report their results for the R\'enyi entropies,
\begingroup
\allowdisplaybreaks
\begin{align}
S^{(n)}(A(B)) &= \Big(1+\frac{1}{n}\Big) \frac{\pi c}{48}\frac{\ell_{1(2)}+\ell_3+\ell_4}\eps  + \log S_{00}\nn\\
&\hspace{0.5cm}+ \frac{1}{1-n}\log\sum_a \vert\psi_a\vert^{2n}(\S_{a0})^{2(1-n)}\,,\\
S^{(n)}(A\cup B) &= \Big(1+\frac{1}{n}\Big) \frac{\pi c}{48}\frac{\ell_1+\ell_2}\eps  + 2\log S_{00}\,,\;\;\lb{RECcontr}
\end{align}
and for the (R\'enyi) mutual information,
\begin{align}
&I^{(n)}(A:B) \notag\\
&\hspace{-0.cm} =\Big(1+\frac{1}{n}\Big) \frac{\pi c}{24}\frac{\ell_3+\ell_4}\eps + \frac{2}{1-n}\log\sum_a \vert\psi_a\vert^{2n} (\S_{a0})^{2(1-n)},\nn\\
&I(A:B) \notag\\ \lb{MITcC}
&\hspace{0.cm}= \frac{\pi c}{12}\frac{\ell_3+\ell_4}\eps +4\sum_a \vert\psi_a\vert^{2} \log\S_{a0}-2\sum_a \vert\psi_a\vert^{2} \log\vert\psi_a\vert^{2}.\nn\\[-.1\baselineskip]
\end{align}
\endgroup

\onecolumngrid
\section{Surgery manifolds} \lb{Apdx2}

We gather here the figures related to the calculation of reflected entropy using the surgery method discussed in Section \ref{sec3}.
\bigskip
\bigskip

\begin{figure*}[h]
\centering
\includegraphics[scale=0.72]{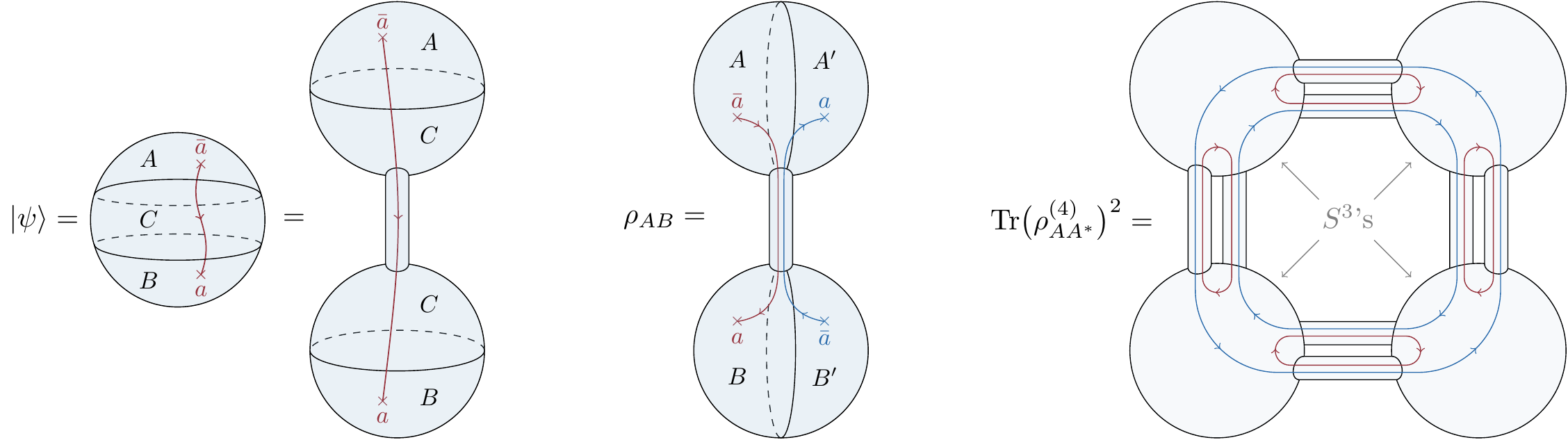}
\caption{Surgery: Two disjoint regions on the 2-sphere. The wave function $\vert \psi\ra$ is deformed into two 3-balls joined by a tube. The reduced density matrix $\rho_{AB}$ is obtained by tracing out the region $C$. After surgery, the manifold corresponding to $\tr(\rho_{AA^*}^{(m)})^n$ is displayed for $m=4$ and $n=2$. It is composed of $S^3$'s connected by tubes along $S^2$'s.}
   \label{fig4}
\end{figure*}
   \begin{figure*}[h]
      \centering
      \includegraphics[width=\textwidth]{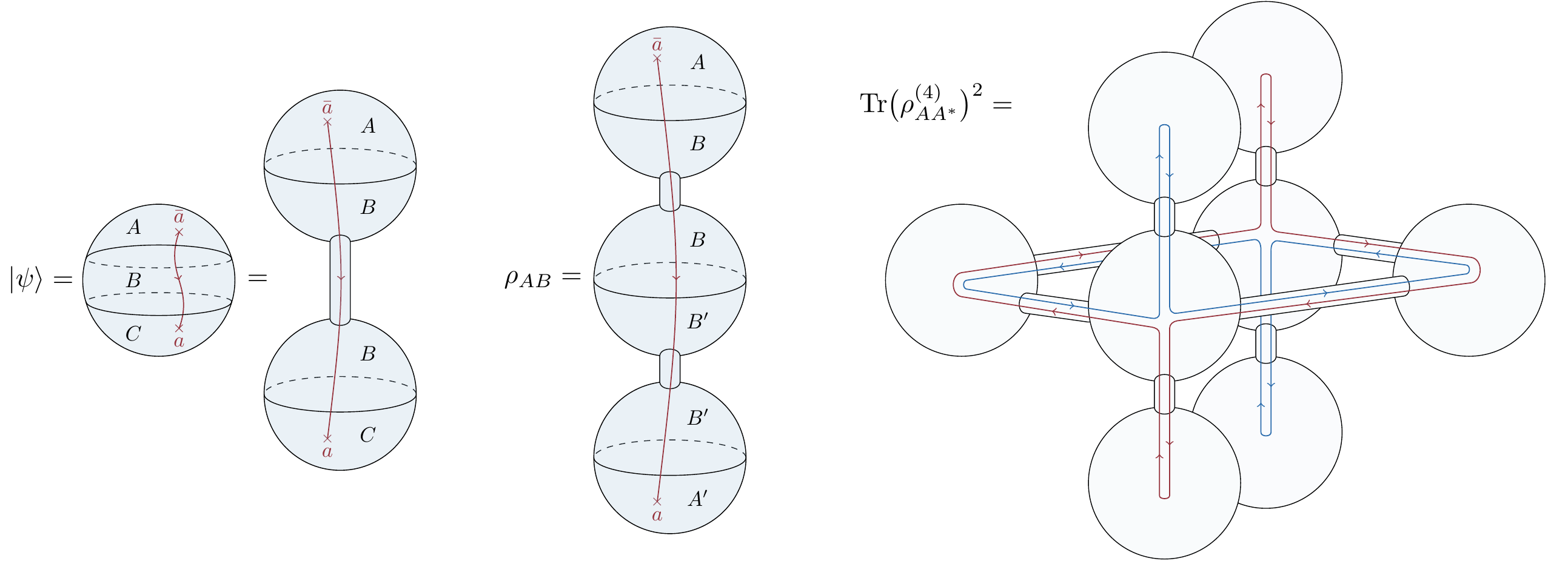}
      \caption{Surgery: Two adjacent regions on the 2-sphere. The wave function $\vert \psi\ra$ is deformed into two 3-balls joined by a tube. The reduced density matrix $\rho_{AB}$ is obtained as three 3-balls joined along two tubes. The manifold corresponding to $\tr(\rho_{AA^*}^{(m)})^n$ is displayed for $m=4$ and $n=2$, and it is composed of $S^3$'s connected by tubes along $S^2$'s.}
      \label{fig5}
   \end{figure*}
   \begin{figure*}[h]
      \centering
      \includegraphics[width=\textwidth]{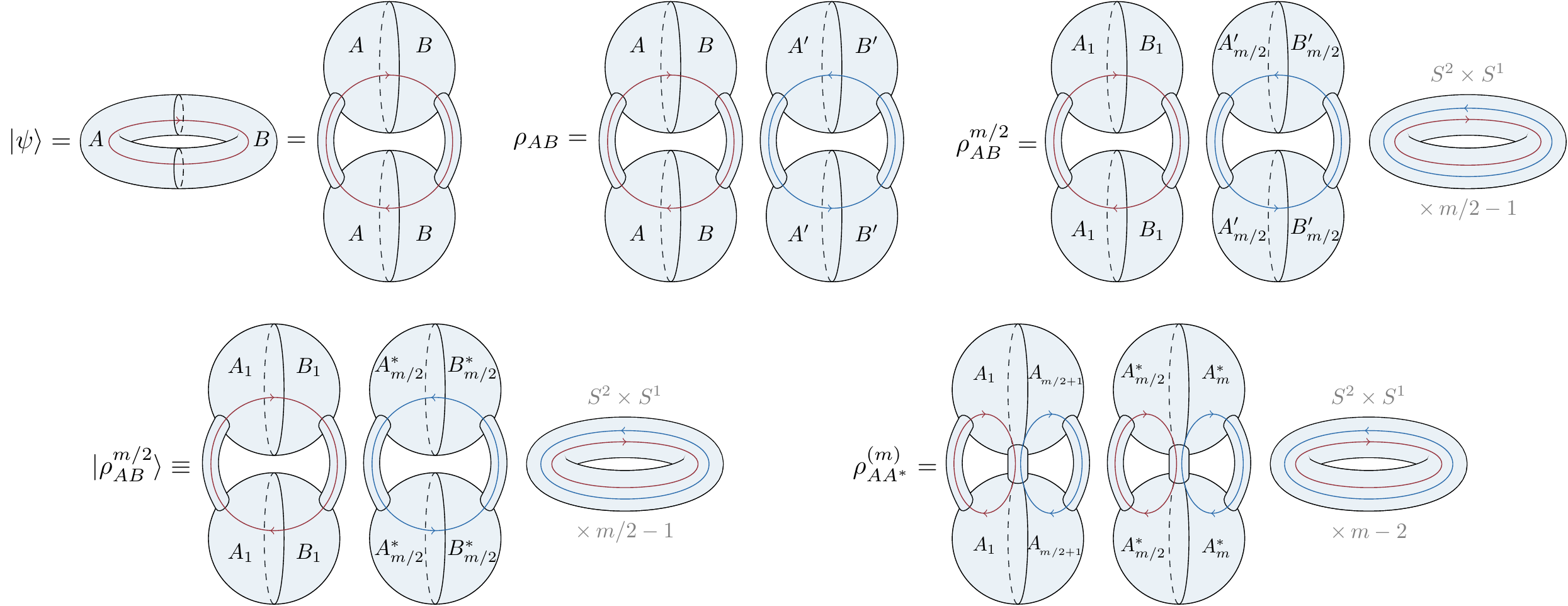}
      \caption{Surgery: Pure bipartite state on the 2-torus. The wave function $\vert \psi\ra$ is deformed into two 3-balls connected by two tubes. The gluing of $m/2$ density matrices results in two pairs of 3-balls joined by two tubes and $m/2-1$ $S^2\times S^1$'s. The canonical purification $\vert \rho_{AB}^{m/2}\ra$ is obtained by interpreting the second pair of 3-balls in $\rho_{AB}^{m/2}$ as living on $\H_{A^*}\otimes\H_{B^*}$. Finally, from $\vert\rho_{AB}^{(m)}\ra$ one gets the reduced density matrix $\rho_{AA^*}^{(m)}$ by tracing out the regions $B$ and $B^*$.}
      \label{fig6}
   \end{figure*}
\begin{figure*}[h]
\centering
\includegraphics[scale=0.77]{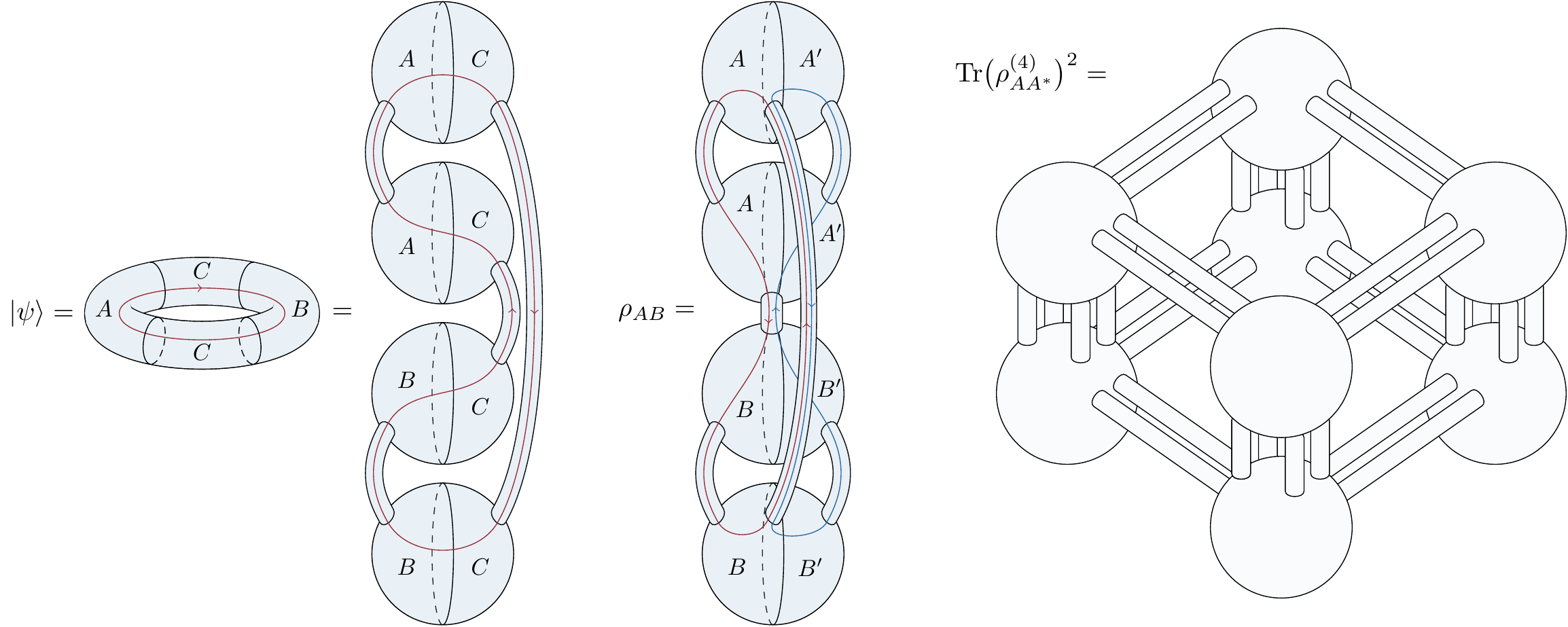}
\caption{Surgery: Two disjoint non-contractible regions on the 2-torus. The wave function $\vert \psi\ra$ is deformed into four 3-balls joined by three tubes, with a Wilson loop threading through all the regions. The reduced density matrix $\rho_{AB}$ is obtained as for 3-balls joined by six tubes. The manifold corresponding to $\tr(\rho_{AA^*}^{(m)})^n$ is displayed for $m=4$ and $n=2$, and it is composed of $S^3$'s connected by tubes along $S^2$'s. We do not show the Wilson lines since their paths would render the figure illegible.}
   \label{fig7}
\end{figure*}
   \begin{figure*}[h]
      \centering
      \includegraphics[scale=0.82]{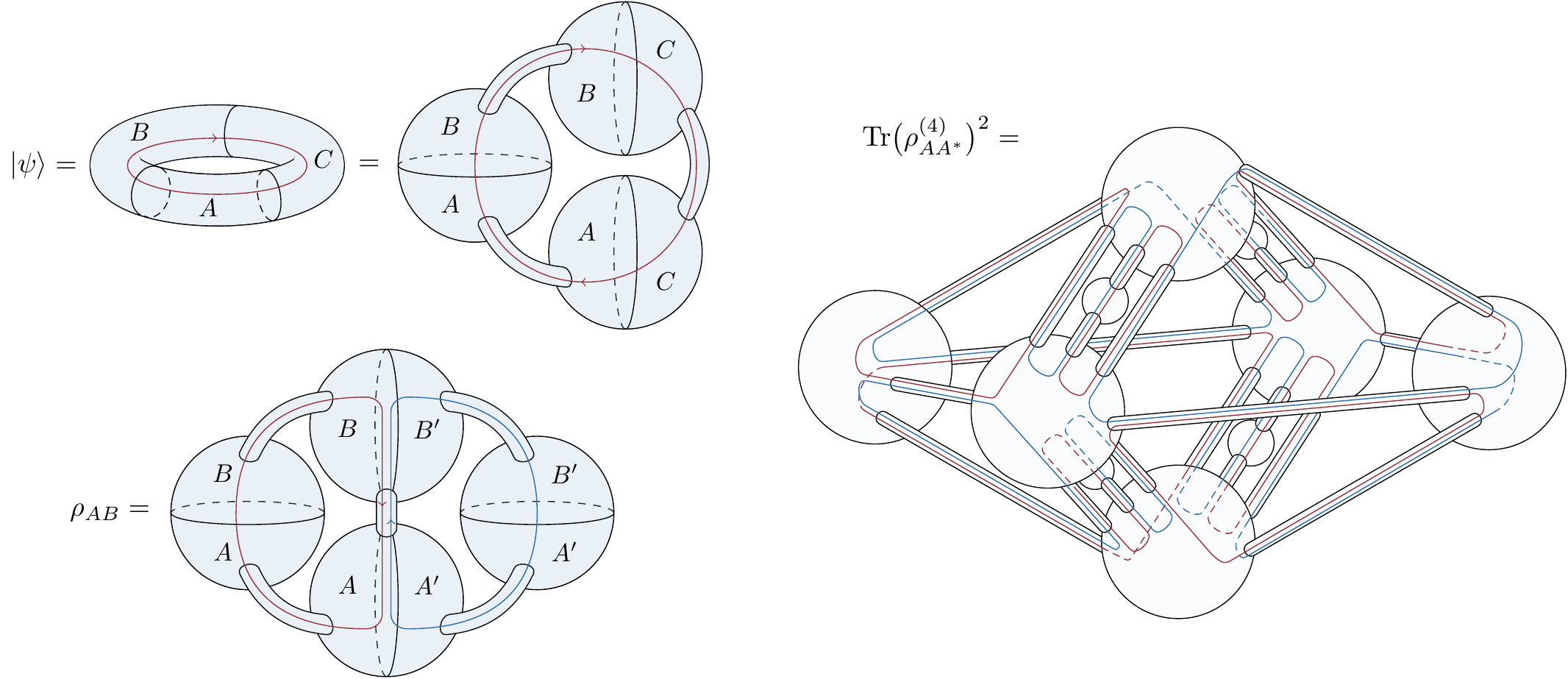}
      \caption{Surgery: Two adjacent non-contractible regions on the 2-torus. The wave function $\vert \psi\ra$ is deformed into three 3-balls joined by three tubes, with a Wilson loop threading through all the regions. The reduced density matrix $\rho_{AB}$ is obtained as four 3-balls connected by four tubes. The manifold corresponding to $\tr(\rho_{AA^*}^{(m)})^n$ is displayed for $m=4$ and $n=2$, and it is composed of $S^3$'s connected by tubes along $S^2$'s.}
      \label{fig9}
   \end{figure*}
\begin{figure*}[h]
\centering
\includegraphics[scale=0.81]{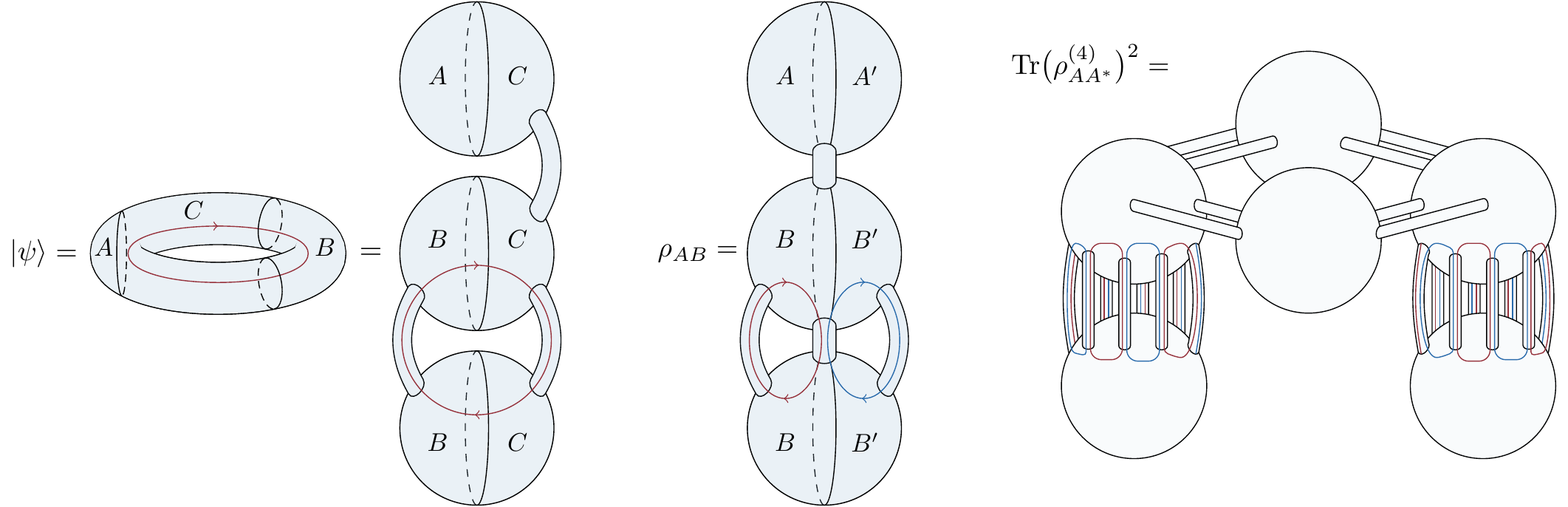}
\caption{Surgery: Two disjoint regions on the 2-torus with contractible $A$ and non-contractible $B$. The wave function $\vert \psi\ra$ is deformed into three 3-balls joined by three tubes, with a Wilson loop threading through the regions $B$ and $C$ only. The reduced density matrix $\rho_{AB}$ is obtained as three 3-balls joined by four tubes. The manifold corresponding to $\tr(\rho_{AA^*}^{(m)})^n$ is displayed for $m=4$ and $n=2$, and it is composed of $S^3$'s connected by tubes along $S^2$'s.}
   \label{fig8}
\end{figure*}
\begin{figure*}[h]
	\centering
	\includegraphics[width=\textwidth]{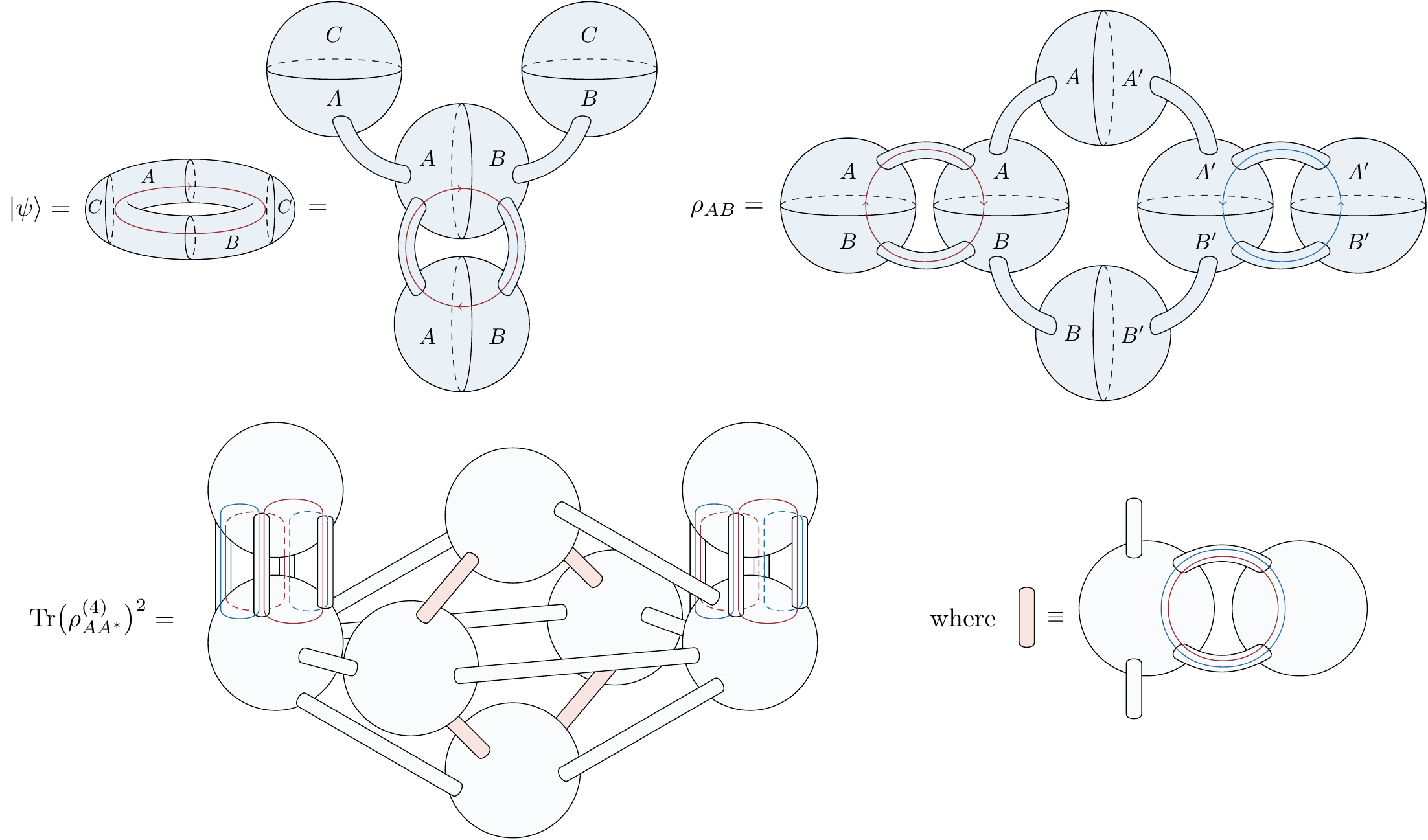}
	\caption{Surgery: Two adjacent non-contractible regions with contractible $C$. The wave function $\vert \psi\ra$ is deformed into four 3-balls joined by four tubes, with a Wilson loop threading through the regions $A$ and $B$ only. The reduced density matrix $\rho_{AB}$ is obtained as six 3-balls connected by eight tubes. The manifold corresponding to $\tr(\rho_{AA^*}^{(m)})^n$ is displayed for $m=4$ and $n=2$, and it is composed of $S^3$'s connected by tubes along $S^2$'s.}
	\label{fig10}
\end{figure*}

\FloatBarrier
\twocolumngrid



\providecommand{\href}[2]{#2}\begingroup\raggedright\endgroup


\end{document}